\documentclass[twocolumn,english,aps,prd,floatfix,nofootinbib,superscriptaddress,preprintnumbers]{revtex4-2}
\usepackage[T1]{fontenc}
\usepackage[utf8]{inputenc}
\usepackage{xcolor}
\usepackage{calc}
\usepackage{mathrsfs}
\usepackage{amsmath}
\usepackage{graphicx}
\usepackage{microtype}

\makeatletter

\usepackage{mathrsfs}
\usepackage{bm}

\usepackage{babel}
\usepackage{xcolor}

\newcommand{\tpi}{\tilde{\pi}}
\newcommand{\tf}{\tilde{f}}
\newcommand{\txi}{\tilde{\xi}}
\newcommand{\tV}{\tilde{V}}
\newcommand{\tphi}{\tilde{\phi}}
\newcommand{\tVp}{\tilde{V}_{,\tphi}}

\newcommand{\ba}{\begin{eqnarray}}
\newcommand{\ea}{\end{eqnarray}}
\newcommand{\be}{\begin{equation}}
\newcommand{\ee}{\end{equation}}

\def\Mpl{M_{\rm P}}

\newcommand{\D}{{\rm d}}

\makeatother

\usepackage{babel}
\begin{document}

\preprint{YITP-26-97}

\title{Minimally modified gravity with Laplacian auxiliary constraints and
an inflationary realization}
\author{Jakkrit Sangtawee}
\affiliation{The Institute for Fundamental Study (The Tah Poe Academia Institute),
Naresuan University, Phitsanulok 65000, Thailand}
\author{Antonio De Felice}
\affiliation{Center for Gravitational Physics and Quantum Information, Yukawa Institute
for Theoretical Physics, Kyoto University, 606-8502 Kyoto, Japan}
\author{Khamphee Karwan}
\affiliation{The Institute for Fundamental Study (The Tah Poe Academia Institute),
Naresuan University, Phitsanulok 65000, Thailand}
\begin{abstract}
We construct a minimally modified gravity theory that propagates only
two tensorial gravitational degrees of freedom around a spatially
flat Friedmann--Lemaître--Robertson--Walker (FLRW) background and
admits a predictive cosmological perturbation theory. We first show
that, in the original four-constraint construction, the homogeneous
values of the Lagrange multipliers are not fully determined and nevertheless
enter the quadratic tensor action, thereby obstructing cosmological
predictivity. We remove this ambiguity by coupling the auxiliary constraints
to spatial Laplacians of the multipliers. The multiplier sector then
drops out of the homogeneous background equations while continuing
to constrain the inhomogeneous scalar sector. The tensor dispersion
relation generically contains both $k^{2}$ and $k^{4}$ contributions,
whereas no propagating gravitational vector or scalar mode is present
on the generic branch for which the constraint reduction is nondegenerate. We subsequently
study a subclass whose gravitational Hamiltonian density is proportional
to the lapse and contains cubic momentum invariants. In this subclass the
$k^{4}$ tensor term vanishes and
the lapse can be absorbed into a time redefinition at the background
level. After minimally coupling a canonical inflaton, the only propagating
scalar mode is the inflaton fluctuation, with unit sound speed. For
a quadratic potential, departures from general relativity shift the
scalar spectral index, while a tensor speed $c_{T}>1$ suppresses
the tensor-to-scalar ratio according to $r=16\epsilon_{s}/c_{T}$.
The parameter regions compatible with the observational bounds adopted
in this work require a superluminal tensor speed; however, very large
$c_{T}$ simultaneously reduces the tensor kinetic coefficient and may
lower the perturbative cutoff, although determining the actual
strong-coupling scale requires a nonlinear analysis.
\end{abstract}
\maketitle

\section{Introduction}

General relativity (GR), supplemented by a cosmological constant and
the standard matter components, provides an exceptionally successful
description of the observed Universe. The resulting $\Lambda$CDM
model accounts for the cosmic microwave background, large-scale structure,
and the late-time accelerated expansion with a remarkably small number
of parameters \cite{Kosowsky:2002zt,Durrer:2015lza}. Its theoretical
interpretation nevertheless remains incomplete, most notably because
the observed vacuum-energy scale is extremely small compared with
naive quantum-field-theory estimates \cite{Sahni:2002kh,Weinberg:1988cp}.
At earlier times, an accelerated phase of expansion offers a compelling
solution to the horizon and flatness problems and provides a mechanism
for generating primordial perturbations \cite{Starobinsky:1980te,Guth:1980zm}.
These considerations motivate extensions of the gravitational sector
that remain close to GR while allowing qualitatively new cosmological
dynamics.

A generic modification of gravity introduces additional propagating
degrees of freedom. Such modes are not necessarily pathological, but
they enlarge the space of stability conditions, can mediate additional
forces, and may complicate the coupling to matter. Tests of GR with
gravitational waves have made this issue especially sharp \cite{LIGOScientific:2016lio},
while the multimessenger observation of GW170817 and GRB~170817A
constrains the low-redshift tensor propagation speed to be extremely
close to the speed of light \cite{Monitor:2017mdv}. It is therefore
useful to identify modifications of gravity that alter the background
and tensor dynamics without adding local gravitational scalar or vector
modes.

Theories with only two tensorial gravitational degrees of freedom
evade one or more assumptions of Lovelock's theorem \cite{Lovelock:1970zsf,Lovelock:1972vz}.
An early example is the Cuscuton, in which a non-dynamical scalar
modifies the gravitational equations without introducing an additional
local mode \cite{Afshordi:2006ad}. A systematic route is provided
by minimally modified gravity (MMG), formulated with only spatial
diffeomorphism invariance and an appropriate constraint structure
\cite{Lin:2017oow,Mukohyama:2019unx}. Matter coupling is subtle in
constructions where the elimination of the unwanted mode relies on
a first-class Hamiltonian constraint, because adding matter can change
its class \cite{Carballo-Rubio:2018czn}. This problem can be avoided
by introducing a gauge-fixing or auxiliary constraint from the outset,
so that the relevant constraints are already second class before matter
is coupled \cite{Aoki:2018zcv,Aoki:2020oqc}. Cosmological applications
of matter-coupled MMG have been explored, for example, in Ref.~\cite{Aoki:2020oqc,Sangtawee:2021mhz,DeFelice:2022mcd,DeFelice:2023bwq}.

A Hamiltonian MMG construction with four auxiliary constraints was
proposed in Ref.~\cite{Yao:2023qjd}. As we show below, its direct
cosmological implementation has a specific obstruction: on an FLRW
background, the homogeneous Lagrange multipliers are not fully fixed,
yet they enter the coefficients governing tensor perturbations. Consequently,
the propagation of gravitational waves cannot be predicted from the
background solution and the physical initial conditions alone. This
is not an instability by itself, but it prevents the theory from defining
a closed cosmological model.

In this work we modify the auxiliary sector so that the constraints
are multiplied by spatial Laplacians of the Lagrange multipliers.
For inhomogeneous Fourier modes, the resulting elliptic equations
impose the same local constraints under standard boundary conditions,
whereas the homogeneous multiplier contribution vanishes identically.
We derive the background equations and the quadratic actions for tensor,
vector, and scalar perturbations. We then minimally couple a canonical
scalar field and identify a subclass in which the gravitational Hamiltonian
density is proportional to the lapse and the homogeneous equations are
invariant under time reparametrizations. Finally, we investigate
slow-roll inflation in a cubic-momentum toy model and determine its
scalar and tensor spectra.

The paper is organized as follows. In Sec.~\ref{sec2} we review
the original four-constraint construction and exhibit its cosmological
obstruction. Section~\ref{sec:new-theory} introduces the Laplacian
auxiliary constraints and studies the gravitational perturbations.
In Sec.~\ref{sec:scalar-field} we couple a canonical scalar field
and discuss the role of the lapse. Section~\ref{sec:inflation} develops
the inflationary model and its observational predictions. We summarize
our results and limitations in Sec.~\ref{sec:conclusions}. 

\section{Original four-constraint MMG construction}

\label{sec2}

General classes of MMG can be formulated in Arnowitt--Deser--Misner
(ADM) variables. The lapse, shift, and induced spatial metric are
denoted by $N$, $N^{i}$, and $\gamma_{ij}$, with conjugate momenta
$\pi_{N}$, $\pi_{i}$, and $\pi^{ij}$, respectively. The covariant
derivative compatible with $\gamma_{ij}$ is denoted by $D_{i}$.
In vacuum, auxiliary constraints can be introduced through the total
Hamiltonian \cite{Yao:2023qjd} 
\begin{equation}
H_{\text{T}}=\int\D^{3}x\left(\mathscr{H}+\mu_{I}\mathcal{S}^{I}+N^{i}\mathcal{H}_{i}+\lambda^{i}\pi_{i}\right),\label{Int_HT_2}
\end{equation}
where the index $I$ labels the auxiliary constraints $\mathcal{S}^{I}$
and $\mu_{I}$ are their Lagrange multipliers. Both $\mathscr{H}$
and $\mathcal{S}^{I}$ are generic functions of $\left(N,\pi_{N},\gamma_{ij},\pi^{ij},D_{i}\right)$.
The multipliers $N^{i}$ and $\lambda^{i}$ enforce the momentum-sector
constraints 
\begin{equation}
\mathcal{H}_{i}\approx0_{i},\quad\pi_{i}\approx0_{i},\label{Int_3d_diff}
\end{equation}
where $\approx$ denotes weakly vanishing. The constraints $\mathcal{H}_{i}$
generate spatial diffeomorphisms. The extended momentum constraints
$\mathcal{H}_{i}$ are defined as \cite{Gao:2014fra,Mukohyama:2015gia,Saitou:2016lvb}
\begin{eqnarray}
\mathcal{H}_{i}\equiv &  & -2\sqrt{\gamma}D_{j}\frac{\pi^{j}_{i}}{\sqrt{\gamma}}+\pi_{N}D_{i}N\nonumber \\
 &  & +\pi_{j}D_{i}N^{j}+\sqrt{\gamma}D_{j}\frac{\pi_{i}N^{j}}{\sqrt{\gamma}}\approx0_{i}.\label{Int_Hi}
\end{eqnarray}
The first term in Eq.~\eqref{Int_Hi} is the usual metric momentum
constraint. When smeared with an arbitrary spatial vector
$\xi^{i}(\boldsymbol{x})$, it generates spatial diffeomorphisms of
the canonical pair $(\gamma_{ij},\pi^{ij})$ through their spatial
Lie derivatives. A spatial translation is only the special case of
a constant $\xi^{i}$ in Cartesian coordinates. Since $N$ and $N^{i}$
are also included as canonical coordinates in the present phase space,
the generator must additionally transform them as a spatial scalar
and a spatial vector, respectively. The terms proportional to $\pi_N$
and $\pi_i$ provide precisely this extension. Indeed, defining
$\mathcal{H}[\boldsymbol{\xi}]\equiv\int {\rm d}^{3}x\,
\xi^{i}\mathcal{H}_{i}$, one obtains
\begin{equation}
\{N,\mathcal{H}[\boldsymbol{\xi}]\}
=\mathcal{L}_{\boldsymbol{\xi}}N,\qquad
\{N^{i},\mathcal{H}[\boldsymbol{\xi}]\}
=\xi^{j}D_{j}N^{i}-N^{j}D_{j}\xi^{i}.
\end{equation}
For a time-dependent spatial-diffeomorphism parameter, the complete
gauge generator also contains the primary constraints $\pi_i$, with
coefficients involving $\dot{\xi}^{i}$ and shift-dependent terms.
Thus $(\pi_i,\mathcal{H}_i)$ form the first-class set associated with
the three spatial-diffeomorphism gauge functions.

In the class considered below, $\pi_N\approx0$ and $\pi_i\approx0$
are included among the primary constraints. The added terms therefore
vanish on the primary-constraint surface. Consequently, the extended
and unextended momentum constraints agree weakly: the extension neither
changes the physical constraint surface nor introduces a new symmetry.
Rather, it gives the correct first-class representation, on the full
ADM phase space, of the spatial-diffeomorphism invariance already
assumed in constructing the theory.
The six first-class constraints $(\pi_{i},\mathcal{H}_{i})$ remove
twelve phase-space dimensions, corresponding to six configuration-space
degrees of freedom. Before imposing the auxiliary constraints, the
remaining system therefore contains four configuration-space degrees
of freedom. Reducing this number to the two tensorial gravitational
modes requires either four second-class constraints or one first-class
plus two second-class constraints. We focus on the former option,
for which the elimination of the unwanted modes does not rely on a
first-class Hamiltonian constraint and is consequently robust under
minimal matter coupling. Since the lapse carries no time derivative,
its conjugate momentum can be included among the auxiliary constraints:
\begin{equation}
\mathcal{S}^{4}=\pi_{N}\approx0\,.
\end{equation}
The total Hamiltonian in Eq.~\eqref{Int_HT_2} then becomes 
\begin{align}
H_{\text{T}} & =\int\D^{3}x\left[\mathscr{H}(N,\gamma_{ij},\pi^{ij},D_{i})+\mu_{I}\mathcal{S}^{I}+N^{i}\mathcal{H}_{i}\right.\nonumber \\
 & +\left.\lambda^{i}\pi_{i}+\lambda_{N}\pi_{N}\right],
\end{align}
where now $I=1,2,3$, while the fourth multiplier has been renamed
$\lambda_{N}$. To classify the possible forms of $\mathcal{S}^{I}$,
we restrict $\mathscr{H}$ to be an arbitrary function of $\left(N,\gamma_{ij},\pi^{ij},R_{ij}\right)$
\cite{Yao:2020tur}, where $R_{ij}$ is the Ricci tensor of the spatial
hypersurface. The generalized Cayley--Hamilton theorem limits the
independent scalar invariants constructed from $\pi^{ij}$ and $R_{ij}$
\cite{Mertzios:1986cht}. Accordingly, $\mathcal{S}^{I}$ and $\mathscr{H}$
can be expressed as generic functions of $(N,\mathcal{R}^{I},\Pi^{I},\mathcal{Q}^{I})$,
where 
\begin{eqnarray}
\mathcal{R}^{I} & \equiv & \left\{ R^{i}_{i},R^{i}_{j}R^{j}_{i},R^{i}_{j}R^{j}_{k}R^{k}_{i}\right\} \,,\\
\Pi^{I} & \equiv & \left\{ \tpi^{i}_{i},\tpi^{i}_{j}\tpi^{j}_{i},\tpi^{i}_{j}\tpi^{j}_{k}\tpi^{k}_{i}\right\} \,,\\
\mathcal{Q}^{I} & \equiv & \left\{ R^{i}_{j}\tpi^{j}_{i},R^{i}_{j}\tpi^{j}_{k}\tpi^{k}_{i},R^{i}_{j}R^{j}_{k}\tpi^{k}_{i}\right\} \,,
\end{eqnarray}
where the tensor density has been replaced by $\tilde{\pi}^{ij}\equiv\pi^{ij}/\sqrt{\gamma}$.
Reference~\cite{Yao:2023qjd} adopted the constraints 
\begin{equation}
\mathcal{S}^{I}=\mathcal{Q}^{I}-\mathcal{P}^{I}(N)\approx0\,,
\end{equation}
where $\mathcal{P}^{I}$ are generic functions of $N$. We begin with
this explicit realization and then explain why its multiplier coupling
should be modified for cosmological applications. 
Rank condition for the minimal theory.
The statement that no additional minimalizing condition is required
in the four-constraint construction presupposes that the four auxiliary
constraints form an independent second-class set \cite{Yao:2023qjd}.
For the present realization, the corresponding local non-degeneracy
condition can be stated explicitly. For a fixed nonzero Fourier mode,
with delta distributions and momentum labels suppressed, define
$p_I\equiv\mathcal{P}^{I}{}_{,N}$,
$C_{IJ}\equiv\{\mathcal{S}^{I},\mathcal{S}^{J}\}$, and
$\Phi_A=(\pi_N,\mathcal{S}^{1},\mathcal{S}^{2},\mathcal{S}^{3})$.
Using the convention
$\{N(\boldsymbol{x}),\pi_N(\boldsymbol{y})\}
=\delta^{3}(\boldsymbol{x}-\boldsymbol{y})$, the Dirac matrix,
evaluated on the constraint surface, is
\begin{equation}
\Delta_{AB}\equiv\{\Phi_A,\Phi_B\}\big|_{\Phi\approx0}=
\begin{pmatrix}
0&p_J\\
-p_I&C_{IJ}
\end{pmatrix}.
\end{equation}
A three-dimensional antisymmetric matrix has rank zero or two. On
the generic rank-two branch, let $v^I$ span its one-dimensional kernel,
$C_{IJ}v^J=0$. The four auxiliary constraints are then second class
precisely when
\begin{equation}
\operatorname{rank}\Delta=4
\qquad\Longleftrightarrow\qquad
p_Iv^I\neq0\,.
\end{equation}
In that case they remove the two unwanted configuration-space degrees
of freedom. Since the functions $\mathcal{P}^{I}(N)$ are freely
specifiable, this condition can be imposed as a non-degeneracy
requirement on the branch of interest. It must nevertheless be checked
throughout the relevant region of phase space and for every nonzero
mode. Zeros of $p_Iv^I$, as well as configurations for which
$\operatorname{rank}C<2$, define degenerate branches on which the
Dirac algorithm must be continued. In particular, the rank can drop
on highly symmetric backgrounds; the primary Dirac matrix alone then
does not determine the perturbative degree-of-freedom count, which we
verify directly for flat FLRW below.

This constraint-algebra condition is logically independent of
$\mathscr{H}_{,NN}$. The latter appears as the coefficient of $\dot N$
in the time derivative of the homogeneous lapse constraint and determines
whether the corresponding consistency condition fixes the lapse evolution.
Hence $\mathscr{H}_{,NN}=0$ does not by itself spoil the second-class
structure, nor does it by itself make the lapse a gauge function.
The absorption of $N$ into a redefinition of time is instead a
property of the lapse-proportional homogeneous subclass studied in
Sec.~\ref{sec:scalar-field}.
The total Hamiltonian
density is 
\begin{align}
\mathscr{H}_{{\rm T}}={} & \sqrt{\gamma}\tilde{\mathscr{H}}^{{\rm CH}}(N,\Pi^{I},\mathcal{R}^{I})-2\sqrt{\gamma}N_{i}D_{j}\tilde{\pi}^{ij}\nonumber \\
 & +\pi_{N}N_{i}\gamma^{ij}D_{j}N+\sqrt{\gamma}\lambda_{N}\tpi_{N}\nonumber \\
 & +\sqrt{\gamma}\mu_{I}(\mathcal{Q}^{I}-\mathcal{P}^{I})\,.\label{HT}
\end{align}
where $\tilde{\mathscr{H}}^{{\rm CH}}\equiv\mathscr{H}/\sqrt{\gamma}$
and $\tilde{\pi}_{N}\equiv\pi_{N}/\sqrt{\gamma}$. As shown below,
the direct multiplier coupling in Eq.~\eqref{HT} creates a predictivity
problem on an FLRW background. The corresponding Lagrangian is 
\begin{align}
L & =\int\D^{3}x\sqrt{\gamma}\,\Bigl[\dot{\gamma}_{ab}\tilde{\pi}^{ab}-\bigl(\tilde{\mathscr{H}}^{{\rm CH}}(N,\Pi^{I},\mathcal{R}^{I})-2N_{i}D_{j}\tilde{\pi}^{ij}\nonumber \\
 & +\tilde{\pi}_{N}N_{i}\gamma^{ij}D_{j}N+\lambda_{N}\,\tilde{\pi}_{N}+\mu_{I}\,(\mathcal{Q}^{I}-\mathcal{P}^{I})\bigr)\Bigr].
\end{align}
In terms of the extrinsic curvature $K_{ij}\equiv(\dot{\gamma}_{ij}-D_{i}N_{j}-D_{j}N_{i})/(2N)$,
the Lagrangian becomes 
\begin{align}
L=\int\D^{3}x\sqrt{\gamma}\Bigl[ & 2NK_{ij}\tilde{\pi}^{ji}-\tilde{\mathscr{H}}^{{\rm CH}}(N,\Pi^{I},\mathcal{R}^{I})\nonumber \\
 & -\mu_{I}(\mathcal{Q}^{I}-\mathcal{P}^{I})\Bigr],\label{lag-ori}
\end{align}
where we have also performed integration by parts for the term $-2N_{i}D_{j}\tilde{\pi}^{ij}$.
Varying the action $\mathcal{S}=\int\D t\,L$ with respect to $\tilde{\pi}^{ij}$
gives 
\begin{equation}
2NK_{ij}-\frac{\delta\tilde{\mathscr{H}}^{{\rm CH}}}{\delta\tilde{\pi}^{ij}}-\mu_{I}\,\frac{\delta\mathcal{Q}^{I}}{\delta\tilde{\pi}^{ij}}=0\,.
\end{equation}
This equation is intended to determine $\tilde{\pi}^{ij}$ algebraically
in terms of $K_{ij}$ and the remaining fields. Such an inversion
is generally local and may define several branches. In GR, it reproduces
the Einstein--Hilbert action in ADM form.

\subsection{Cosmological obstruction on an FLRW background}

To test the cosmological viability of the theory defined by Eq.~\eqref{lag-ori},
we first study tensor perturbations around a spatially flat FLRW background.
We write 
\begin{equation}
\gamma_{ij}=a(t)^{2}\,(\delta_{ij}+h_{ij})\,,\label{gam-tensor}
\end{equation}
where $h_{ij}$ is transverse and traceless, $\delta^{jl}\partial_{l}h_{ij}=0$
and $\delta^{ij}h_{ij}=0$. The momentum variable is decomposed as
\begin{equation}
\tilde{\pi}^{ij}=\frac{f(t)}{a^{2}}\left[\delta^{ij}+\frac{1}{\sqrt{2}}\,\delta\pi^{ij}\right]\,.\label{dpi-tensor}
\end{equation}
The remaining fields have no tensor components and are therefore purely
homogeneous: 
\begin{align}
N_{i} & =0\,,\quad N=N(t)\,,\\
\mu_{1} & =\mu_{1}(t)\,,\mu_{2}=\mu_{2}(t)\,,\quad\mu_{3}=\mu_{3}(t)\,.
\end{align}

The homogeneous dynamics follows from the Lagrangian density 
\begin{align}
\mathcal{L} & =a^{3}\left[6\frac{\dot{a}}{a}f-\tilde{\mathscr{H}}^{{\rm CH}}(N,3f,3f^{2},3f^{3},0,0,0)\right.\nonumber \\
 & +\mu_{1}\mathcal{P}^{1}(N)+\mu_{2}\mathcal{P}^{2}(N)+\mu_{3}\mathcal{P}^{3}(N)\Bigr],\label{lag}
\end{align}
where 
\begin{equation}
\tilde{\mathscr{H}}^{{\rm CH}}=\tilde{\mathscr{H}}^{{\rm CH}}(N,3f,3f^{2},3f^{3},0,0,0)\,.
\end{equation}
The Lagrangian in Eq.~\eqref{lag} provides the constraint equations
associated with the fields $\mu_{I}$ as 
\begin{align}
a^{3}\mathcal{P}^{1}(N)=0\,,\quad a^{3}\mathcal{P}^{2}(N)=0\,,\quad a^{3}\mathcal{P}^{3}(N)=0\,.
\end{align}
For $a>0$, these equations require $N$ to lie in the common zero
set of the three functions $\mathcal{P}^{I}$. If their common roots
are isolated, the lapse is fixed to a constant value $N=N_{0}$, with
\begin{equation}
\mathcal{P}^{1}(N_{0})=\mathcal{P}^{2}(N_{0})=\mathcal{P}^{3}(N_{0})=0\,.
\end{equation}
Thus an FLRW solution exists only if the three functions admit a common
root (or share a continuous zero set). Moreover, fixing $N=N_{0}$
is not a choice of time coordinate: it is imposed dynamically by the
theory.

The remaining background equations follow from variation with respect
to $a$, $f$, and $N$. Variation with respect to $a$ gives 
\begin{equation}
-6a^{2}{{\dot{f}}}-3a^{2}\tilde{\mathscr{H}}^{{\rm CH}}=0\,,
\end{equation}
where we have already imposed $\mathcal{P}^{1}(N_{0})=\mathcal{P}^{2}(N_{0})=\mathcal{P}^{3}(N_{0})=0$.
Variation with respect to $f$ gives 
\begin{equation}
-9a^{3}f^{2}\,\tilde{\mathscr{H}}^{{\rm CH}}_{,\Pi^{3}}-6a^{3}f\,\tilde{\mathscr{H}}^{{\rm CH}}_{,\Pi^{2}}-3a^{3}\,\tilde{\mathscr{H}}^{{\rm CH}}_{,\Pi^{1}}+6a^{2}{{\dot{a}}}=0\,,\label{adot-ori}
\end{equation}
where a subscript ${},X$ denotes partial differentiation with respect
to $X$. The above equation can be rewritten as 
\begin{equation}
H\equiv\frac{{{\dot{a}}}}{Na}=\frac{1}{2N}\,\tilde{\mathscr{H}}^{{\rm CH}}_{,\Pi^{1}}+\frac{f}{N}\,\tilde{\mathscr{H}}^{{\rm CH}}_{,\Pi^{2}}+\frac{3f^{2}}{2N}\,\tilde{\mathscr{H}}^{{\rm CH}}_{,\Pi^{3}}\,.
\end{equation}
This relation determines $f$ locally in terms of $a$, $\dot{a}$,
and $N_{0}$, provided the relevant branch is invertible. Finally,
variation with respect to $N$ gives 
\begin{align}
0 & =-a^{3}\tilde{\mathscr{H}}^{{\rm CH}}_{,N}+a^{3}\mu_{1}\,\left.\mathcal{P}^{1}{}_{,N}\right|_{N=N_{0}}+a^{3}\mu_{2}\,\left.\mathcal{P}^{2}{}_{,N}\right|_{N=N_{0}}\nonumber \\
 & +a^{3}\mu_{3}\,\left.\mathcal{P}^{3}{}_{,N}\right|_{N=N_{0}}\,.
\end{align}
In general, the derivatives $\mathcal{P}^{I}{}_{,N}|_{N=N_{0}}$ do
not vanish. This single equation then fixes at most one linear combination
of the three multipliers, leaving two combinations arbitrary. If all
$\mathcal{P}^{I}{}_{,N}|_{N=N_{0}}$ vanish, the equation instead
becomes an algebraic relation for $f$ (and hence for $H$ through
the momentum equation), analogous to a first Friedmann equation, while
all three multipliers $\mu_{I}$ remain undetermined on the homogeneous
background. Consequently, any observable that depends on these multipliers---such
as the tensor propagation coefficients below---cannot be predicted
from the cosmological initial data. The tensor action is obtained
by expanding Eq.~\eqref{lag-ori} to second order in the transverse-traceless
perturbation and using Eq.~\eqref{adot-ori} to eliminate $\dot{a}$.
We expand $h_{ij}$ and $\delta\pi_{ij}$ in polarization tensors
$\epsilon^{m}_{ij}$ as 
\begin{eqnarray}
h_{ij} & = & \frac{1}{(2\pi)^{3}}\int d^{3}k\sum_{m}\epsilon^{m}_{ij}(k)h^{m}(t)e^{i{\bf {k}\cdot{\bf {x}}}}\,,\label{hm}\\
\delta\pi_{ij} & = & \frac{1}{(2\pi)^{3}}\int d^{3}k\sum_{m}\epsilon^{m}_{ij}(k)\delta\pi^{m}(t)e^{i{\bf {k}\cdot{\bf {x}}}}\,,\label{pm}
\end{eqnarray}
where $m=+,\times$ labels the two tensor polarizations and $\delta^{ij}\epsilon^{m}_{ij}=k^{i}\epsilon^{m}_{ij}=0$.

Varying the perturbed action with respect to $\delta\pi^{m}$ yields
the constraint equation 
\begin{align}
\delta\pi^{m}={} & \frac{1}{4a^{2}f\left(3f\tilde{\mathscr{H}}^{{\rm CH}}_{,\Pi^{3}}+\tilde{\mathscr{H}}^{{\rm CH}}_{,\Pi^{2}}\right)}\nonumber \\
 & \quad\times\Bigl\{2a^{2}\dot{h}_{m}-h_{m}\left(12a^{2}f^{2}\tilde{\mathscr{H}}^{{\rm CH}}_{,\Pi^{3}}-2f\mu_{2}k^{2}\right.\nonumber \\
 & \hspace{8em}\left.-4a^{2}f\tilde{\mathscr{H}}^{{\rm CH}}_{,\Pi^{2}}-\mu_{1}k^{2}\right)\Bigr\}\,.
\end{align}

Substituting this constraint into the quadratic action gives 
\begin{align}
S_{T}=\sum_{m}\int{\rm d}t\,{\rm d}^{3}k\,Na^{3}\Biggl[ & \tilde{Q}_{T}\frac{\dot{h}^{2}_{m}}{N^{2}}-\tilde{D}_{T}\frac{k^{4}}{a^{4}}h^{2}_{m}\nonumber \\
 & -\tilde{C}_{T}\frac{k^{2}}{a^{2}}h^{2}_{m}\Biggr]\,,
\end{align}
or, equivalently, 
\begin{align}
\mathcal{S}_{T}=\int\D t\D^{3}x\,Na^{3}\Biggl[ & \tilde{Q}_{T}\frac{\dot{h}^{2}_{m}}{N^{2}}-\frac{\tilde{D}_{T}}{a^{4}}(\partial^{2}h_{m})^{2}\nonumber \\
 & -\frac{\tilde{C}_{T}}{a^{2}}\delta^{ij}(\partial_{i}h_{m})(\partial_{j}h_{m})\Biggr],
\end{align}
where $\partial^{2}\equiv\delta^{ij}\partial_{i}\partial_{j}$, and 
\begin{widetext}
\begin{eqnarray}
\tilde{Q}_{T} & = & \frac{N}{12f\tilde{\mathscr{H}}^{{\rm CH}}_{,\Pi^{3}}+4\tilde{\mathscr{H}}^{{\rm CH}}_{,\Pi^{2}}}\,,\\
\tilde{D}_{T} & = & \frac{-\left(f\mu_{3}+\tilde{\mathscr{H}}^{{\rm CH}}_{,\mathcal{R}^{3}}\right)\tilde{\mathscr{H}}^{{\rm CH}}_{,\Pi^{2}}-3f\left(f\mu_{3}+\tilde{\mathscr{H}}^{{\rm CH}}_{,\mathcal{R}^{3}}\right)\tilde{\mathscr{H}}^{{\rm CH}}_{,\Pi^{3}}+\left(\mu_{2}f+\frac{\mu_{1}}{2}\right)^{2}}{4N\left(3f\tilde{\mathscr{H}}^{{\rm CH}}_{,\Pi^{3}}+\tilde{\mathscr{H}}^{{\rm CH}}_{,\Pi^{2}}\right)}\,\\
\tilde{C}_{T} & = & \frac{1}{16\left(3f\tilde{\mathscr{H}}^{{\rm CH}}_{,\Pi^{3}}+\tilde{\mathscr{H}}^{{\rm CH}}_{,\Pi^{2}}\right)^{2}N}\Bigg[\left(8f^{2}\mu_{2}+6\mu_{1}f+4\tilde{\mathscr{H}}^{{\rm CH}}_{,\mathcal{R}^{2}}\right)\left(\tilde{\mathscr{H}}^{{\rm CH}}_{,\Pi^{2}}\right)^{2}+\left(54f^{4}\mu_{2}+45\mu_{1}f^{3}+36f^{2}\tilde{\mathscr{H}}^{{\rm CH}}_{,\mathcal{R}^{2}}\right)\left(\tilde{\mathscr{H}}^{{\rm CH}}_{,\Pi^{3}}\right)^{2}\nonumber \\
 &  & +\left[(42f^{3}\mu_{2}+33\mu_{1}f^{2}+24f\tilde{\mathscr{H}}^{{\rm CH}}_{,\mathcal{R}^{2}})\tilde{\mathscr{H}}^{{\rm CH}}_{,\Pi^{3}}-2\tilde{\mathscr{H}}^{{\rm CH}}\mu_{2}+\left(2\mu_{2}f+\mu_{1}\right)\tilde{\mathscr{H}}^{{\rm CH}}_{,\Pi^{1}}+4f\dot{\mu}_{2}+2\dot{\mu}_{1}\right]\tilde{\mathscr{H}}^{{\rm CH}}_{,\Pi^{2}}\nonumber \\
 &  & +\left\{ 3\tilde{\mathscr{H}}^{{\rm CH}}\mu_{1}+6f\left[\left(\mu_{2}f+\frac{\mu_{1}}{2}\right)\tilde{\mathscr{H}}^{{\rm CH}}_{,\Pi^{1}}+2f\dot{\mu}_{2}+\dot{\mu}_{1}\right]\right\} \tilde{\mathscr{H}}^{{\rm CH}}_{,\Pi^{3}}\nonumber \\
 &  & -4\left[\left(-\frac{27}{2}\tilde{\mathscr{H}}^{{\rm CH}}_{,\Pi^{3}\Pi^{3}}f^{3}-\frac{27}{2}\tilde{\mathscr{H}}^{{\rm CH}}_{,\Pi^{2}\Pi^{3}}f^{2}-\frac{9}{2}\tilde{\mathscr{H}}^{{\rm CH}}_{,\Pi^{1}\Pi^{3}}f-3\tilde{\mathscr{H}}^{{\rm CH}}_{,\Pi^{2}\Pi^{2}}f-\frac{3}{2}\tilde{\mathscr{H}}^{{\rm CH}}_{,\Pi^{1}\Pi^{2}}\right)\tilde{\mathscr{H}}^{{\rm CH}}\right.\nonumber \\
 &  & +\dot{N}\left(3\tilde{\mathscr{H}}^{{\rm CH}}_{,N\Pi^{3}}f+\tilde{\mathscr{H}}^{{\rm CH}}_{,N\Pi^{2}}\right)\Bigr]\left(\mu_{2}f+\frac{\mu_{1}}{2}\right)\Bigg]\,.
\end{eqnarray}
\end{widetext}

The tensor action depends explicitly on the homogeneous values of
$\mu_{I}$ and on their time derivatives. Since the background equations
do not determine these functions, the tensor propagation coefficients
are not fixed by the cosmological solution. The original construction
is therefore not predictive on an FLRW background without additional
conditions on the multiplier sector.

\section{Laplacian auxiliary-constraint theory}

\label{sec:new-theory}

To remove the undetermined homogeneous multipliers while retaining
the local auxiliary constraints, we modify the total Hamiltonian density
as 
\begin{align}
\mathscr{H}_{{\rm tot}} & =\sqrt{\gamma}\tilde{\mathscr{H}}^{{\rm CH}}(N,\Pi^{I},\mathcal{R}^{I})-2\sqrt{\gamma}N_{i}D_{j}\tilde{\pi}^{ij}\nonumber \\
 & +\sqrt{\gamma}\tilde{\pi}_{N}N_{i}\gamma^{ij}D_{j}N+\sqrt{\gamma}\lambda_{N}\,\tilde{\pi}_{N}\nonumber \\
 & +\sqrt{\gamma}\,(\gamma^{ij}D_{i}D_{j}\mu_{I})\,(\mathcal{Q}^{I}-\mathcal{P}^{I})\,.\label{hnew}
\end{align}
The phase-space dependence of $\mathcal{Q}^{I}$ and $\mathcal{P}^{I}$
is the same as in the original construction, with $\mathcal{P}^{I}=\mathcal{P}^{I}(N)$.
Variation with respect to $\mu_{I}$ now gives the elliptic equations
\begin{equation}
D^{2}\!\left(\mathcal{Q}^{I}-\mathcal{P}^{I}\right)=0,\qquad D^{2}\equiv\gamma^{ij}D_{i}D_{j}.\label{eq:elliptic-constraints}
\end{equation}
For Fourier modes with $k\neq0$, and under boundary conditions that
exclude non-trivial harmonic functions, Eq.~\eqref{eq:elliptic-constraints}
is locally equivalent to $\mathcal{Q}^{I}-\mathcal{P}^{I}=0$. Its
homogeneous kernel is deliberately left unconstrained. Decomposing
$\mu_{I}=\bar{\mu}_{I}(t)+\delta\mu_{I}$, one has $D^{2}\bar{\mu}_{I}=0$
on FLRW, and hence the homogeneous multipliers drop out of the background
equations. The scalar perturbations $\delta\mu_{I}$ continue to impose
constraints on inhomogeneous modes, while they do not contribute to
the linear tensor sector. Global zero modes can depend on the spatial
topology and boundary conditions and should be treated separately.
The phase-space Lagrangian of the modified theory is 
\begin{align}
L={} & \int\D^{3}x\sqrt{\gamma}\,\Biggl\{\dot{\gamma}_{ij}\tilde{\pi}^{ij}+\dot{N}\tilde{\pi}_{N}\nonumber \\
 & -\Bigl[\tilde{\mathscr{H}}^{{\rm CH}}(N,\Pi^{I},\mathcal{R}^{I})-2N_{i}D_{j}\tilde{\pi}^{ij}\nonumber \\
 & \quad+\tilde{\pi}_{N}N_{i}\gamma^{ij}D_{j}N+\lambda_{N}\tilde{\pi}_{N}\nonumber \\
 & \quad+(\gamma^{ij}D_{i}D_{j}\mu_{I})(\mathcal{Q}^{I}-\mathcal{P}^{I})\Bigr]\Biggr\},
\end{align}
After eliminating the primary lapse momentum and integrating the shift
term by parts, this becomes 
\begin{align}
L & =\int\D^{3}x\sqrt{\gamma}\left[2NK_{ij}\tilde{\pi}^{ji}-\tilde{\mathscr{H}}^{{\rm CH}}(N,\Pi^{I},\mathcal{R}^{I})\right.\nonumber \\
 & -\left.(\gamma^{ij}D_{i}D_{j}\mu_{I})\,(\mathcal{Q}^{I}-\mathcal{P}^{I})\right].\label{lag-new}
\end{align}

In this case, the background Lagrangian density is given by 
\begin{equation}
\mathcal{L}=a^{3}\left[6\frac{\dot{a}}{a}f-\tilde{\mathscr{H}}^{{\rm CH}}\left(N,3f,3f^{2},3f^{3},0,0,0\right)\right].
\end{equation}
The corresponding background equations are 
\begin{align}
\tilde{\mathscr{H}}^{{\rm CH}}+2\dot{f} & =0\,,\label{eq:2ndEinstein}\\
-3f^{2}\,\tilde{\mathscr{H}}^{{\rm CH}}_{,\Pi^{3}}-2f\,\tilde{\mathscr{H}}^{{\rm CH}}_{,\Pi^{2}}-\tilde{\mathscr{H}}^{{\rm CH}}_{,\Pi^{1}}+2\,\frac{\dot{a}}{a} & =0\,,\label{eq:f_momentum}\\
\tilde{\mathscr{H}}^{{\rm CH}}_{,N} & =0\,.\label{eq:Fried1}
\end{align}
All quantities are evaluated on the homogeneous and isotropic background,
\[
\Pi^{1}=3f,\qquad\Pi^{2}=3f^{2},\qquad\Pi^{3}=3f^{3},
\]
while the remaining arguments of $\tilde{\mathscr{H}}^{{\rm CH}}$
vanish.

The lapse equation~\eqref{eq:Fried1} is algebraic rather than
evolutionary. Since it depends only on $N$ and $f$ on the background,
\[
\tilde{\mathscr{H}}^{{\rm CH}}_{,N}
=\tilde{\mathscr{H}}^{{\rm CH}}_{,N}(N,f),
\]
it constrains these two variables. Its preservation in time requires
\begin{equation}
0=\frac{{\rm d}}{{\rm d}t}\tilde{\mathscr{H}}^{{\rm CH}}_{,N}
=\tilde{\mathscr{H}}^{{\rm CH}}_{,NN}\dot N
+\frac{\partial}{\partial f}
\left(\tilde{\mathscr{H}}^{{\rm CH}}_{,N}\right)\dot f\,.
\end{equation}
Thus, when $\tilde{\mathscr{H}}^{{\rm CH}}_{,NN}\neq0$, this
consistency condition determines $\dot N$ once $\dot f$ is known.
It does not introduce an additional equation or degree of freedom;
it only propagates the original algebraic constraint along the background
solution. If the lapse Hessian vanishes, the consistency condition loses
the term proportional to $\dot N$ and the corresponding branch must be
analyzed separately.

Moreover, Eq.~\eqref{eq:f_momentum} determines how the auxiliary
quantity $f$ is tied to the background expansion. Provided this equation
is locally invertible, it allows one to write
\[
f=\mathcal{F}(H,N),\qquad H\equiv\frac{\dot{a}}{Na}.
\]
Inserting this result into Eq.~\eqref{eq:Fried1} yields an algebraic
equation involving only $H$ and $N$. This relation plays the role
of the first Friedmann equation in the present Hamiltonian formulation.

\subsection{Tensor perturbations}

To study gravitational waves, the metric tensor $\gamma_{ij}$ and
the momentum $\tilde{\pi}^{ij}$ are expanded around the FLRW background
according to Eqs.~\eqref{gam-tensor} and \eqref{dpi-tensor}. Expanding
Eq.~\eqref{lag-new} to second order in tensor perturbations gives
\begin{equation}
\mathcal{S}_{T}=\int\D t\,\D^{3}x\,Na^{3}\mathcal{L}_{T}\,,\label{act-ten}
\end{equation}
where 
\begin{align}
\mathcal{L}_{T} & =\frac{f}{N}\dot{h}_{ij}\delta\pi^{ij}-\frac{f^{2}}{N}(\tilde{\mathscr{H}}^{{\rm CH}}_{,\Pi^{2}}+3f\tilde{\mathscr{H}}^{{\rm CH}}_{,\Pi^{3}})\delta\pi_{ij}\delta\pi^{ij}\nonumber \\
 & -\frac{2f^{2}}{N}(\tilde{\mathscr{H}}^{{\rm CH}}_{,\Pi^{2}}+3f\tilde{\mathscr{H}}^{{\rm CH}}_{,\Pi^{3}})h_{ij}\delta\pi^{ij}+\frac{\tilde{\mathscr{H}}^{{\rm CH}}_{,\mathcal{R}^{1}}}{4a^{2}N}\partial_{k}h_{ij}\partial^{k}h^{ij}\nonumber \\
 & -\frac{\tilde{\mathscr{H}}^{{\rm CH}}_{,\mathcal{R}^{2}}}{4a^{4}N}\partial^{2}h_{ij}\partial^{2}h^{ij}\,.\label{lag-tpt}
\end{align}
Equation~\eqref{eq:f_momentum} has been used to eliminate $\dot{a}$,
whereas $\dot{f}$ has been removed by means of Eq.~\eqref{eq:2ndEinstein}.
Since the lapse has not been fixed, $t$ denotes a general time coordinate
rather than necessarily cosmic time. After transforming to Fourier
space, variation with respect to $\delta\pi^{m}$ yields 
\begin{equation}
\delta\pi^{m}=\frac{\dot{h}_{m}-2fh_{m}\left(3f\tilde{\mathscr{H}}^{{\rm CH}}_{,\Pi^{3}}+\tilde{\mathscr{H}}^{{\rm CH}}_{,\Pi^{2}}\right)}{2f\left(3f\tilde{\mathscr{H}}^{{\rm CH}}_{,\Pi^{3}}+\tilde{\mathscr{H}}^{{\rm CH}}_{,\Pi^{2}}\right)}\,.\label{delpij}
\end{equation}
Substituting Eq.~\eqref{delpij} back into the quadratic action gives
\begin{align}
\mathcal{S}_{T}=\sum_{m}\int\D t\,\D^{3}k\,Na^{3}\Biggl[ & Q_{T}\frac{\dot{h}^{2}_{m}}{N^{2}}-D_{T}\frac{k^{4}}{a^{4}}h^{2}_{m}\nonumber \\
 & -C_{T}\frac{k^{2}}{a^{2}}h^{2}_{m}\Biggr],\label{act-tensor}
\end{align}
where 
\begin{align}
Q_{T} & =\frac{N}{4\,\bigl(3f\tilde{\mathscr{H}}^{{\rm CH}}_{,\Pi^{3}}+\tilde{\mathscr{H}}^{{\rm CH}}_{,\Pi^{2}}\bigr)}\,,\label{cq-tensor}\\
D_{T} & =\frac{\tilde{\mathscr{H}}^{{\rm CH}}_{,\mathcal{R}^{2}}}{4N}\,,\\
C_{T} & =-\frac{\tilde{\mathscr{H}}^{{\rm CH}}_{,\mathcal{R}^{1}}}{4N}\,.\label{cc-tensor}
\end{align}
The coefficients depend only on the Hamiltonian and on background
quantities determined by the background equations. Unlike in the original
construction of Sec.~\ref{sec2}, no undetermined homogeneous multiplier
enters the tensor action.

For GR, 
\begin{equation}
\tilde{\mathscr{H}}^{{\rm CH}}_{{\rm GR}}=-\frac{\Mpl^{2}}{2}\,N\mathcal{R}^{1}-\frac{N\,(\Pi^{1})^{2}}{\Mpl^{2}}+\frac{2N\Pi^{2}}{\Mpl^{2}}\,,
\end{equation}
so that 
\begin{equation}
Q_{T{\rm GR}}=\frac{\Mpl^{2}}{8}\,,\qquad D_{T{\rm GR}}=0\,,\qquad C_{T{\rm GR}}=\frac{\Mpl^{2}}{8}\,,
\end{equation}
which reproduces the standard luminal tensor propagation of GR.

For the general theory, the tensor dispersion relation is 
\begin{equation}
\omega^{2}=\frac{D_{T}}{Q_{T}}\,\frac{k^{4}}{a^{4}}+\frac{C_{T}}{Q_{T}}\,\frac{k^{2}}{a^{2}}\,.\label{eq:dispersion}
\end{equation}
The no-ghost condition is $Q_{T}>0$. If the $k^{4}$ term is present,
high-momentum stability additionally requires $D_{T}>0$; if this
term is absent, gradient stability instead requires $C_{T}/Q_{T}>0$.
If the effective theory remains valid up to the frequencies probed
by late-time multimessenger observations, compatibility with an approximately
luminal, frequency-independent propagation speed requires 
\begin{equation}
D_{T}=0\,,\qquad C_{T}=Q_{T}\,.
\end{equation}
The reason is that the multimessenger bound concerns low-redshift
astrophysical waves with physical momenta far above cosmological scales.
If $D_{T}$ were nonzero, the $k^{4}$ contribution would dominate
at sufficiently large $k/a$ and generate a strongly scale-dependent
propagation speed. Within the regime of validity of the effective
theory, imposing $D_{T}=0$ and $C_{T}=Q_{T}$ is equivalent to 
\begin{equation}
\tilde{\mathscr{H}}^{{\rm CH}}_{,\mathcal{R}^{2}}=0\,,\qquad-\frac{\tilde{\mathscr{H}}^{{\rm CH}}_{,\mathcal{R}^{1}}}{N}=\frac{N}{3f\tilde{\mathscr{H}}^{{\rm CH}}_{,\Pi^{3}}+\tilde{\mathscr{H}}^{{\rm CH}}_{,\Pi^{2}}}\,.
\end{equation}
For an inflationary effective description that subsequently approaches
GR, these late-time restrictions need only hold after the transition
to the low-energy regime.

\subsection{Vector perturbations}

In vacuum, no vector matter source is present. To verify that the
gravitational sector does not propagate a vector mode, we decompose
\begin{equation}
\gamma_{ij}=a^{2}\delta_{ij}+a\,(\partial_{i}C_{j}+\partial_{j}C_{i})\,.
\end{equation}
The vector part of $\tilde{\pi}^{ij}$ is parameterized by 
\begin{equation}
\tilde{\pi}^{ij}=\frac{f}{a^{2}}\,\delta^{ij}+\frac{f}{a^{3}}\,\delta^{im}\delta^{jn}\,(\partial_{m}P_{n}+\partial_{n}P_{m})\,.
\end{equation}
The transverse part of the shift is written as 
\begin{equation}
N_{i}=a\,N^{V}_{i}\,.
\end{equation}
All vector perturbations are transverse; for example, $\partial^{i}C_{i}=0$.
Since the lapse and the multipliers are scalars, they remain purely
homogeneous in the vector sector: $N=N(t)$ and $\mu_{I}=\mu_{I}(t)$.

Using spatial diffeomorphism invariance, we choose the gauge $C_{i}=0$.
Expanding to quadratic order and transforming to Fourier space gives
\begin{align}
S_{V}=\int\D t\,\D^{3}k\,a\Bigl[ & -2k^{2}f\,\delta^{ij}P_{i}N^{V}_{j}\nonumber \\
 & -2k^{2}f^{2}\left(\tilde{\mathscr{H}}^{{\rm CH}}_{,\Pi^{2}}+3f\tilde{\mathscr{H}}^{{\rm CH}}_{,\Pi^{3}}\right)\delta^{ij}P_{i}P_{j}\Bigr].
\end{align}
Variation with respect to $N^{V}_{i}$ imposes $P_{i}=0$ for $k\neq0$
and $f\neq0$. The quadratic vector action then vanishes, so no gravitational
vector mode propagates.

\subsection{Scalar perturbations}

For scalar perturbations, we parameterize the spatial metric as 
\begin{align}
\gamma_{ij}\,\D x^{i}\otimes\D x^{j}={} & [a^{2}(1+2\zeta)\delta_{ij}\nonumber \\
 & +2\partial_{i}\partial_{j}E]\,\D x^{i}\otimes\D x^{j}\,.\label{eq:gamma_pert_scal}
\end{align}
The scalar part of the momentum is parameterized by 
\begin{align}
\tilde{\pi}^{ij}\partial_{i}\otimes\partial_{j}={} & \biggl[\frac{f}{a^{2}}(1+2\zeta_{\pi})\delta^{ij}\nonumber \\
 & +\frac{2f}{a^{4}}\delta^{im}\delta^{jn}\partial_{m}\partial_{n}E_{\pi}\biggr]\partial_{i}\otimes\partial_{j}\,.\label{eq:pi_pert_scal}
\end{align}
The remaining scalar perturbations are 
\begin{align}
N & =N(t)\,(1+\alpha)\,,\qquad N_{i}\,\D x^{i}=\partial_{i}\chi\,\D x^{i}\,,\label{eq:shift_pert_scal}\\
\mu_{I} & =\mu_{I}(t)+\delta\mu_{I}\,.
\end{align}
The three $\delta\mu_{I}$ are independent multiplier perturbations.
We use spatial diffeomorphism invariance to choose the gauge $E=0$.
The quadratic scalar action is summarized in Appendix~\ref{appendix1}.
Variation of this action with respect to $\delta\mu_{3}$ yields 
\begin{equation}
\mathcal{P}^{3}{}_{,N}k^{2}\alpha=0\,.
\end{equation}
For the generic branch with $k\neq0$ and $\mathcal{P}^{3}{}_{,N}\neq0$,
this equation gives $\alpha=0$. Degenerate choices of $\mathcal{P}^{3}$
must be analyzed separately. The remaining multiplier perturbations
impose, on the generic branch, 
\begin{equation}
k^{4}\frac{f}{a}\zeta=0\,,
\end{equation}
which gives $\zeta=0$ for $k\neq0$ and $f\neq0$. Variation with
respect to $\chi$ gives 
\begin{equation}
E_{\pi}=\frac{a^{2}}{k^{2}}\,\zeta_{\pi}\,.
\end{equation}
Substituting this relation together with $\alpha=\zeta=0$ into the
quadratic action, the reduced Lagrangian density becomes 
\begin{align}
\mathcal{L} & =-48\zeta^{2}_{\pi}a^{3}f^{2}\Bigl[\frac{3}{2}f^{4}\tilde{\mathscr{H}}^{{\rm CH}}_{,\Pi^{3}\Pi^{3}}+2f^{3}\tilde{\mathscr{H}}^{{\rm CH}}_{,\Pi^{2}\Pi^{3}}+f^{2}\tilde{\mathscr{H}}^{{\rm CH}}_{,\Pi^{1}\Pi^{3}}\nonumber \\
 & +\frac{2}{3}f^{2}\tilde{\mathscr{H}}^{{\rm CH}}_{,\Pi^{2}\Pi^{2}}+\frac{2}{3}f\tilde{\mathscr{H}}^{{\rm CH}}_{,\Pi^{1}\Pi^{2}}+\frac{1}{6}\tilde{\mathscr{H}}^{{\rm CH}}_{,\Pi^{1}\Pi^{1}}+\frac{1}{6}\tilde{\mathscr{H}}^{{\rm CH}}_{,\Pi^{2}}\nonumber \\
 & +\frac{1}{2}f\tilde{\mathscr{H}}^{{\rm CH}}_{,\Pi^{3}}\Bigr]\,.
\end{align}
Provided the displayed algebraic coefficient is nonzero, the remaining
equation gives $\zeta_{\pi}=0$. Thus no inhomogeneous gravitational scalar mode propagates on the
generic branch for which the constraint reduction is nondegenerate.

\section{Coupling a canonical scalar field}

\label{sec:scalar-field}

The vacuum analysis shows that the modified gravitational sector carries
only the two tensorial modes around FLRW. We now minimally couple
a canonical scalar field and use it as an inflaton. Because the unwanted
gravitational modes are removed by auxiliary constraints that are
already second class, adding minimally coupled matter does not rely
on preserving a first-class Hamiltonian constraint and does not change
the local gravitational degree-of-freedom count \cite{Yao:2023qjd}.
We decompose 
\begin{equation}
\phi(t,\mathbf{x})=\bar{\phi}(t)+\delta\phi(t,\mathbf{x}),
\end{equation}
and suppress the bar on the homogeneous field below.

The Lagrangian for the scalar field can be written as 
\begin{equation}
\mathcal{L}_{{\rm m}}=N\sqrt{\gamma}\left\{ \frac{(\dot{\phi}-N^{i}\partial_{i}\phi)^{2}}{2N^{2}}-\frac{1}{2}\gamma^{ij}\partial_{i}\phi\partial_{j}\phi-V(\phi)\right\} .\label{lag-phi}
\end{equation}
The total Lagrangian is the sum of Eqs.~\eqref{lag-new} and \eqref{lag-phi}.
Its background equations are 
\begin{align}
\frac{\dot{a}}{a} & =\frac{3}{2}f^{2}\tilde{\mathscr{H}}^{{\rm CH}}_{,\Pi^{3}}+f\tilde{\mathscr{H}}^{{\rm CH}}_{,\Pi^{2}}+\frac{1}{2}\tilde{\mathscr{H}}^{{\rm CH}}_{,\Pi^{1}}\,,\label{dota}\\
\tilde{\mathscr{H}}^{{\rm CH}}_{,N} & =-V-\frac{\dot{\phi}^{2}}{2N^{2}}\,,\label{eq:lapse_eom}\\
\frac{\dot{f}}{N} & =-\frac{V}{2}-\frac{\tilde{\mathscr{H}}^{{\rm CH}}}{2N}+\frac{\dot{\phi}^{2}}{4N^{2}}\,,\label{dotf}\\
\frac{\ddot{\phi}}{N^{2}} & =-3H\frac{\dot{\phi}}{N}-V_{,\phi}+\frac{\dot{N}}{N^{2}}\frac{\dot{\phi}}{N}\,.\label{ddotphi}
\end{align}
The scalar equation can equivalently be written in the manifestly
covariant-in-time form 
\begin{equation}
\frac{1}{Na^{3}}\frac{{\rm d}}{{\rm d}t}\left(a^{3}\frac{\dot{\phi}}{N}\right)+V_{,\phi}=0.
\end{equation}
It has the standard minimally coupled form; the modification enters
through the gravitational background equations that determine $H$.

We define $H\equiv\dot{a}/(Na)$. In the generic lapse-nonlinear theory,
the lapse appears explicitly in the Hamiltonian constraint and cannot
be removed by a time-coordinate choice. In the subclass whose Hamiltonian
density is proportional to $N$, by contrast, the homogeneous equations
depend on the lapse only through $N\,{\rm d}t$.

\subsection{Lapse Hessian and background time reparametrization}

The lapse remains an auxiliary, non-propagating variable in all cases
discussed below. When background time-reparametrization invariance is
broken, it is no longer an arbitrary gauge function: the algebraic
equation~\eqref{eq:lapse_eom} constrains it together with the remaining
background variables. In the lapse-proportional homogeneous subclass,
by contrast, the same equation constrains reparametrization-invariant
combinations while $N$ reflects the choice of time coordinate.

The terminology in this subsection refers only to the Hessian of the
gravitational Hamiltonian density with respect to the non-dynamical
lapse. It should not be confused with the usual degeneracy of the
velocity Hessian $L_{,\dot q^{A}\dot q^{B}}$, which concerns the
invertibility of the Legendre transformation. In fact, $\dot N$ is
absent from the Lagrangian and hence $L_{,\dot N\dot N}=0$ in all
the cases below. The lapse-Hessian condition is also logically distinct
from the rank degeneracy of the auxiliary-constraint Dirac matrix
discussed in Sec.~\ref{sec2}.

To make the role of the lapse Hessian explicit, define the homogeneous
lapse constraint
\begin{equation}
\mathcal{C}_{N}\equiv
\tilde{\mathscr{H}}^{{\rm CH}}_{,N}
+V+\frac{\dot{\phi}^{2}}{2N^{2}}=0\,.
\end{equation}
Taking its time derivative and using the scalar-field equation
\eqref{ddotphi}, the terms proportional to $V_{,\phi}$ cancel, as do
the matter contributions proportional to $\dot N$. The consistency
condition therefore becomes
\begin{equation}
0=\dot{\mathcal{C}}_{N}
=\tilde{\mathscr{H}}^{{\rm CH}}_{,NN}\dot N
+\frac{\partial}{\partial f}
\left(\tilde{\mathscr{H}}^{{\rm CH}}_{,N}\right)\dot f
-3H\frac{\dot{\phi}^{2}}{N}\,.
\label{eq:lapse-consistency}
\end{equation}
This is not an independent background equation; it is the condition
that propagates the original algebraic constraint $\mathcal{C}_{N}=0$
along the solution.

Let us first consider the case with a nonvanishing lapse Hessian
\begin{equation}
\tilde{\mathscr{H}}^{{\rm CH}}_{,NN}\neq0\,.
\end{equation}
Equation~\eqref{eq:lapse-consistency} then determines $\dot N$.
Using Eqs.~\eqref{dota} and \eqref{dotf}, together with
$\Pi^{1}=3f$, $\Pi^{2}=3f^{2}$, and $\Pi^{3}=3f^{3}$, one obtains
\begin{widetext}
\begin{align}
\dot{N} & =\frac{3\dot{\phi}^{2}}{2\tilde{\mathscr{H}}^{{\rm CH}}_{,NN}N^{2}}\left(\tilde{\mathscr{H}}^{{\rm CH}}_{,\Pi^{1}}+2f\tilde{\mathscr{H}}^{{\rm CH}}_{,\Pi^{2}}+3f^{2}\tilde{\mathscr{H}}^{{\rm CH}}_{,\Pi^{3}}\right)\nonumber \\
 & \quad+\frac{3\left(2V\,N^{2}+2\tilde{\mathscr{H}}^{{\rm CH}}N-\dot{\phi}^{2}\right)\left(3f^{2}\tilde{\mathscr{H}}^{{\rm CH}}_{,N\Pi^{3}}+2f\tilde{\mathscr{H}}^{{\rm CH}}_{,N\Pi^{2}}+\tilde{\mathscr{H}}^{{\rm CH}}_{,N\Pi^{1}}\right)}{4N\tilde{\mathscr{H}}^{{\rm CH}}_{,NN}}\,.\label{eq:dotN-nondegenerate}
\end{align}
\end{widetext}

Equation~\eqref{eq:dotN-nondegenerate} is meaningful only when the
lapse Hessian is nonzero and must be evolved together with
the original constraint~\eqref{eq:lapse_eom}. Initial data are therefore
not arbitrary: one combination of $N$, $f$, $\phi$, and $\dot{\phi}$
is fixed by the constraint.

We now turn to the case with a vanishing lapse Hessian,
\begin{equation}
\tilde{\mathscr{H}}^{{\rm CH}}_{,NN}=0\,.
\end{equation}
If this condition holds identically in a neighborhood of the branch,
the most general lapse dependence is affine,
\begin{equation}
\tilde{\mathscr{H}}^{{\rm CH}}
=N\,\mathcal{F}(\Pi^{I},\mathcal{R}^{I})
+\mathcal{G}(\Pi^{I},\mathcal{R}^{I})\,.
\end{equation}
The vanishing of $\tilde{\mathscr{H}}^{{\rm CH}}_{,NN}$ alone therefore
does not make the lapse a gauge function and does not ensure that it
can be absorbed into a time redefinition. Rather, the differentiated
constraint~\eqref{eq:lapse-consistency} loses the term that would
determine $\dot N$ and becomes an additional relation among the other
background variables. If the lapse Hessian vanishes only at an isolated
locus, the evolution equation for $N$ becomes singular there and that
locus must be analyzed separately. We henceforth specialize to the
stronger lapse-proportional subclass
\begin{equation}
\tilde{\mathscr{H}}^{{\rm CH}}=N\,\mathcal{F}(\Pi^{I},\mathcal{R}^{I})\,.
\end{equation}
On the homogeneous and isotropic background one has $\mathcal{R}^{I}=0$,
and hence 
\[
\mathcal{F}=\mathcal{F}(f)\,.
\]
The background equations then reduce to 
\begin{align}
\frac{\dot{a}}{Na} & =\frac{3}{2}f^{2}\mathcal{F}_{,\Pi^{3}}+f\mathcal{F}_{,\Pi^{2}}+\frac{1}{2}\mathcal{F}_{,\Pi^{1}}\,,\\
\mathcal{F} & =-V-\frac{\dot{\phi}^{2}}{2N^{2}}\,,\label{eq:lapse_eom-1}\\
\frac{\dot{f}}{N} & =-\frac{V}{2}-\frac{\mathcal{F}}{2}+\frac{\dot{\phi}^{2}}{4N^{2}}\,,\\
\frac{\ddot{\phi}}{N^{2}} & =-\frac{3\dot{\phi}}{2N}\left(3\mathcal{F}_{,\Pi^{3}}f^{2}+2\mathcal{F}_{,\Pi^{2}}f+\mathcal{F}_{,\Pi^{1}}\right)+\frac{\dot{N}}{N^{2}}\frac{\dot{\phi}}{N}-V_{,\phi}\,.
\end{align}
In this lapse-proportional subclass, the lapse appears in the background
equations only through the proper-time combination $N(t)\,{\rm d}t$. This is
manifest for the first-derivative terms. The scalar-field equation
has the same property, since the combination involving $\ddot{\phi}$
and $\dot{N}$ can be rewritten as 
\begin{equation}
\frac{1}{N}\frac{{\rm d}}{{\rm d}t}\left(\frac{1}{N}\frac{{\rm d}\phi}{{\rm d}t}\right).
\end{equation}
Therefore, at the homogeneous and isotropic level, the lapse can be
completely absorbed into a redefinition of time. This statement applies
only to the background dynamics. Beyond the homogeneous sector, the
lapse perturbation is instead fixed by the constraint structure of
the theory, in such a way that the gravity sector propagates only
tensor degrees of freedom.

It is then convenient to describe the background evolution in terms
of the e-fold variable 
\begin{equation}
\mathcal{N}\equiv\ln\left(\frac{a}{a_{*}}\right),
\end{equation}
for which 
\begin{equation}
\dot{\mathcal{N}}=NH\,.
\end{equation}
Using $\mathcal{N}$ as the time variable, and denoting derivatives
with respect to $\mathcal{N}$ by a prime, the equations become 
\begin{align}
H & =H(\mathcal{N})=\frac{3}{2}f^{2}\mathcal{F}_{,\Pi^{3}}+f\mathcal{F}_{,\Pi^{2}}\nonumber \\
 & \quad+\frac{1}{2}\mathcal{F}_{,\Pi^{1}}\,,\\
\mathcal{F}(f) & =-V-\frac{1}{2}H^{2}(\phi')^{2}\,,\\
Hf' & =-\frac{V}{2}-\frac{\mathcal{F}}{2}+\frac{1}{4}H^{2}(\phi')^{2}\,,\\
H^{2}\phi'' & =-HH'\phi'-3H^{2}\phi'-V_{,\phi}\,.
\end{align}
In this form, the lapse has disappeared from the background equations.
Thus, in the lapse-proportional homogeneous subclass, $N$ does not
represent an independent background variable, but only reflects the
freedom to choose the time coordinate.

\subsection{Reduced scalar action}

Expanding the total Lagrangian, Eqs.~\eqref{lag-new} and \eqref{lag-phi},
to second order gives the scalar action displayed in Eq.~\eqref{act-sc-phi}.
Variation of the perturbed action~\eqref{act-sc-phi} with respect
to $\delta\mu_{3}$ gives the constraint equation 
\begin{equation}
\mathcal{P}^{3}{}_{,N}\,k^{2}\alpha=0\,.
\end{equation}
On the generic branch, $k\neq0$ and $\mathcal{P}^{3}{}_{,N}\neq0$,
so this gives 
\begin{equation}
\alpha=0\,,
\end{equation}
Variation with respect to $\delta\mu_{1}$ and $\delta\mu_{2}$ then
gives 
\begin{equation}
\frac{f}{a}k^{4}\zeta=0\,,
\end{equation}
which implies $\zeta=0$ for $k\neq0$ and $f\neq0$. Variation with
respect to $\chi$ gives 
\begin{equation}
E_{\pi}=\frac{a^{2}}{k^{2}}\zeta_{\pi}-\frac{1}{4f}\frac{a^{2}}{k^{2}}\frac{\dot{\phi}}{N}\delta\phi\,.
\end{equation}

The remaining algebraic equation determines $\zeta_{\pi}$ as 
\begin{widetext}
\begin{equation}
\zeta_{\pi}=-\frac{9\tilde{\mathscr{H}}^{{\rm CH}}_{,\Pi^{3}\Pi^{3}}f^{4}+12\tilde{\mathscr{H}}^{{\rm CH}}_{,\Pi^{2}\Pi^{3}}f^{3}+6\tilde{\mathscr{H}}^{{\rm CH}}_{,\Pi^{1}\Pi^{3}}f^{2}+4\tilde{\mathscr{H}}^{{\rm CH}}_{,\Pi^{2}\Pi^{2}}f^{2}+4\tilde{\mathscr{H}}^{{\rm CH}}_{,\Pi^{1}\Pi^{2}}f+\tilde{\mathscr{H}}^{{\rm CH}}_{,\Pi^{1}\Pi^{1}}}{\frac{3}{2}\tilde{\mathscr{H}}^{{\rm CH}}_{,\Pi^{3}\Pi^{3}}f^{4}+2\tilde{\mathscr{H}}^{{\rm CH}}_{,\Pi^{2}\Pi^{3}}f^{3}+\tilde{\mathscr{H}}^{{\rm CH}}_{,\Pi^{1}\Pi^{3}}f^{2}+\frac{2}{3}\tilde{\mathscr{H}}^{{\rm CH}}_{,\Pi^{2}\Pi^{2}}f^{2}+\frac{1}{2}\tilde{\mathscr{H}}^{{\rm CH}}_{,\Pi^{3}}f+\frac{2}{3}\tilde{\mathscr{H}}^{{\rm CH}}_{,\Pi^{1}\Pi^{2}}f+\frac{1}{6}\tilde{\mathscr{H}}^{{\rm CH}}_{,\Pi^{2}}+\frac{1}{6}\tilde{\mathscr{H}}^{{\rm CH}}_{,\Pi^{1}\Pi^{1}}}\,\frac{1}{48\,f}\,\frac{\dot{\phi}}{N}\,\delta\phi\,.
\end{equation}
\end{widetext}

Substituting the algebraic constraints into the quadratic action gives
\begin{align}
\mathcal{S}=\int{\rm d}^{4}x \frac{Na^{3}}2\!\Biggl[ & \frac{\dot{\delta\phi}^{\,2}}{N^{2}}-\frac{(\partial_{i}\delta\phi)(\partial^{i}\delta\phi)}{a^{2}} 
 -C_{\phi}\,\delta\phi^{2}\Biggr]\!,\label{act-scalar}
\end{align}

where the background-dependent effective mass is 
\begin{widetext}
\begin{align}
C_{\phi} & =V_{,\phi\phi}+\frac{\dot{\phi}^{2}}{N^{2}}\,\frac{D_{\phi}}{E_{\phi}}\left(\frac{3f\tilde{\mathscr{H}}^{{\rm CH}}_{,\Pi^{3}}}{N}+\frac{\tilde{\mathscr{H}}^{{\rm CH}}_{,\Pi^{2}}}{N}\right)\,,\\
D_{\phi} & =27\tilde{\mathscr{H}}^{{\rm CH}}_{,\Pi^{3}\Pi^{3}}f^{4}+36\tilde{\mathscr{H}}^{{\rm CH}}_{,\Pi^{2}\Pi^{3}}f^{3}+18\tilde{\mathscr{H}}^{{\rm CH}}_{,\Pi^{1}\Pi^{3}}f^{2}+12\tilde{\mathscr{H}}^{{\rm CH}}_{,\Pi^{2}\Pi^{2}}f^{2}\nonumber \\
 & \quad+6\tilde{\mathscr{H}}^{{\rm CH}}_{,\Pi^{3}}f+12\tilde{\mathscr{H}}^{{\rm CH}}_{,\Pi^{1}\Pi^{2}}f+2\tilde{\mathscr{H}}^{{\rm CH}}_{,\Pi^{2}}+3\tilde{\mathscr{H}}^{{\rm CH}}_{,\Pi^{1}\Pi^{1}}\,,\\
E_{\phi} & =36\tilde{\mathscr{H}}^{{\rm CH}}_{,\Pi^{3}\Pi^{3}}f^{4}+48\tilde{\mathscr{H}}^{{\rm CH}}_{,\Pi^{2}\Pi^{3}}f^{3}+\left(16\tilde{\mathscr{H}}^{{\rm CH}}_{,\Pi^{2}\Pi^{2}}+24\tilde{\mathscr{H}}^{{\rm CH}}_{,\Pi^{1}\Pi^{3}}\right)f^{2}\nonumber \\
 & \quad+\left(12\tilde{\mathscr{H}}^{{\rm CH}}_{,\Pi^{3}}+16\tilde{\mathscr{H}}^{{\rm CH}}_{,\Pi^{1}\Pi^{2}}\right)f+4\tilde{\mathscr{H}}^{{\rm CH}}_{,\Pi^{2}}+4\tilde{\mathscr{H}}^{{\rm CH}}_{,\Pi^{1}\Pi^{1}}\,.
\end{align}
\end{widetext}

For $\tilde{\mathscr{H}}^{{\rm CH}}=N\mathcal{F}(\Pi^{I},\mathcal{R}^{I})$,
$C_{\phi}$ depends only on the background variables $\phi$, $\phi'$,
and $f$. The kinetic and gradient terms are canonical, so the scalar
no-ghost and gradient conditions are automatically satisfied on this
branch; the sign of $C_{\phi}$ alone is not a stability criterion
during inflation.

\section{Inflationary application}

\label{sec:inflation}

\subsection{Cubic-momentum toy model}

As a concrete inflationary realization, we consider the toy Hamiltonian
\begin{align}
\mathscr{H}^{{\rm CH}} & =\mathscr{H}_{{\rm GR}}+\sqrt{\gamma}N\left[\xi_{1}(\Pi^{1})^{3}+\xi_{2}\Pi^{1}\Pi^{2}+\xi_{3}\Pi^{3}\right]\,,\label{hch-n1}\\
\mathscr{H}_{{\rm GR}} & =\sqrt{\gamma}N\left[-\frac{\Mpl^{2}}{2}\mathcal{R}^{1}-\frac{(\Pi^{1})^{2}}{\Mpl^{2}}+\frac{2\Pi^{2}}{\Mpl^{2}}\right]\,,\\
\mathscr{H}_{{\rm tot}} & =\mathscr{H}^{{\rm CH}}-2\sqrt{\gamma}N_{i}D_{j}\tilde{\pi}^{ij}+\pi_{N}N_{i}\gamma^{ij}D_{j}N\nonumber \\
 & \quad+\lambda_{N}\pi_{N}+\sqrt{\gamma}\,(\gamma^{ij}D_{i}D_{j}\mu_{I})\,(\mathcal{Q}^{I}-\mathcal{P}^{I})\,,
\end{align}
where $\mathscr{H}_{{\rm GR}}$ is the Hamiltonian density for GR.
For this model, the tensor coefficients in Eqs.~\eqref{cq-tensor}--\eqref{cc-tensor}
reduce to 
\begin{eqnarray}
Q_{T} & = & \frac{\Mpl^{2}}{4\left[2+(\xi_{2}+\xi_{3})\,(3f)\Mpl^{2}\right]}\,,\label{qt}\\
C_{T} & = & \frac{\Mpl^{2}}{8}\,,\\
D_{T} & = & 0\,.\label{dt}
\end{eqnarray}
Hence $c^{2}_{T}=C_{T}/Q_{T}=1+\frac{3}{2}f\Mpl^{2}(\xi_{2}+\xi_{3})$,
and tensor modes are luminal on FLRW when $\xi_{2}=-\xi_{3}$. An
illustrative subclass with luminal tensor propagation on FLRW is obtained
by imposing $\xi_{2}=-\xi_{3}$: 
\begin{align}
\mathscr{H}^{{\rm CH}} & =\mathscr{H}_{{\rm GR}}+\sqrt{\gamma}N\left[\xi_{1}(\Pi^{1})^{3}+\xi_{3}(\Pi^{3}-\Pi^{1}\Pi^{2})\right]\,,\\
\mathscr{H}_{{\rm GR}} & =\sqrt{\gamma}N\left[-\frac{\Mpl^{2}}{2}\mathcal{R}^{1}-\frac{(\Pi^{1})^{2}}{\Mpl^{2}}+\frac{2\Pi^{2}}{\Mpl^{2}}\right]\,,\\
\mathscr{H}_{{\rm tot}} & =\mathscr{H}^{{\rm CH}}-2\sqrt{\gamma}N_{i}D_{j}\tilde{\pi}^{ij}+\pi_{N}N_{i}\gamma^{ij}D_{j}N\nonumber \\
 & \quad+\lambda_{N}\pi_{N}+\sqrt{\gamma}\,(D^{2}\mu_{I})\,(\mathcal{Q}^{I}-\mathcal{P}^{I})\,.
\end{align}
For this choice, tensor modes propagate luminally on an FLRW background.
Propagation on less symmetric backgrounds, including compact-object
geometries, can impose additional restrictions.

The scalar effective mass in this toy model is 
\begin{align}
C_{\phi}={} & V_{,\phi\phi}+\frac{\dot{\phi}^{2}}{N^{2}}\frac{-2/\Mpl^{2}+6f(9\xi_{1}+3\xi_{2}+\xi_{3})}{4f(18\xi_{1}+7\xi_{2}+3\xi_{3})}\nonumber \\
 & \quad\times\left(\frac{2}{\Mpl^{2}}+3f\xi_{3}+3f\xi_{2}\right)\,.\label{Cphi}
\end{align}

\subsection{Background evolution}

We now specialize to an inflationary background. For the toy model
in Eq.~\eqref{hch-n1}, 
\begin{equation}
\mathcal{F}=-\frac{\Mpl^{2}}{2}\,\mathcal{R}^{1}-\frac{(\Pi^{1})^{2}}{\Mpl^{2}}+\frac{2\Pi^{2}}{\Mpl^{2}}+\xi_{1}\,(\Pi^{1})^{3}+\xi_{2}\,\Pi^{1}\Pi^{2}+\xi_{3}\,\Pi^{3}.
\end{equation}
The background equations are 
\begin{align}
E & \equiv\frac{H}{H_{i}}=-\tilde{f}+\frac{3{\tilde{f}}^{2}(9\tilde{\xi}_{1}+3\tilde{\xi}_{2}+\tilde{\xi}_{3})}{2}\,,\label{eq1}\\
0 & =\tilde{V}+\frac{E^{2}}{2}\,({\tilde{\phi}'})^{2}-3{\tilde{f}}^{2}+3{\tilde{f}}^{3}\,(9\tilde{\xi}_{1}+3\tilde{\xi}_{2}+\tilde{\xi}_{3})\,,\label{eq2}\\
{\tilde{f}}' & =\frac{E}{2}\,({\tilde{\phi}'})^{2}\,,\label{eq3}\\
\tilde{\phi}'' & =\frac{3\tilde{f}\tilde{\phi}'}{E}-\frac{E'\tilde{\phi}'}{E}-\frac{9\tilde{f}^{2}\tilde{\phi}'}{2E}\,(9\tilde{\xi}_{1}+3\tilde{\xi}_{2}+\tilde{\xi}_{3})-\frac{\tilde{V}_{\tilde{\phi}}}{E^{2}}\,,\label{eq4}
\end{align}
where $H_{i}$ is the initial value of the Hubble parameter and tildes
denote dimensionless variables defined as 
\begin{eqnarray}
\tf & \equiv & \frac{f}{\Mpl^{2}H_{i}}\,,\label{eq:dimensionless-1}\\
\txi_{i} & \equiv & \xi_{i}\Mpl^{4}H_{i}\,,\\
\tV & \equiv & \frac{V}{\Mpl^{2}H^{2}_{i}}\,,\\
\tVp & \equiv & \frac{V_{,\phi}}{\Mpl H^{2}_{i}}\,,\\
\tphi & \equiv & \frac{\phi}{\Mpl}\,.
\end{eqnarray}
It is convenient to define the parameter 
\begin{equation}
\frac{c}{3}\equiv-\left(\txi_{3}+9\txi_{1}+3\txi_{2}\,\right),\label{defC}
\end{equation}
so that Eq.~\eqref{eq2} relates $\tf$ to the scalar-field variables:
\begin{equation}
\frac{c}{3}=\frac{\tV}{3\tf^{3}}+\frac{E^{2}\tphi'{}^{2}}{6\tf^{3}}-\frac{1}{\tf}\,.\label{cSol}
\end{equation}
In terms of the parameter $c$, Eq.~\eqref{eq1} becomes 
\begin{equation}
E=-\frac{1}{2}\tilde{f}(2+c\tilde{f})\,.\label{eOfF}
\end{equation}
Substituting $\tf$ from Eq.~\eqref{cSol} into Eq.~\eqref{eOfF}
and setting $c=0$, we recover the Friedmann equation of GR: 
\begin{equation}
E=\sqrt{\frac{2\tV}{6-\tphi'{}^{2}}}\,.\label{Friedmann}
\end{equation}
Thus $c=0$ gives the Einstein limit of the background dynamics. It
is possible to eliminate the parameters $\txi_{1},\txi_{2}$, and
$\txi_{3}$ from the evolution equations for the background by substituting
$c$ from Eq.~\eqref{cSol} into Eqs.~\eqref{eq1}, \eqref{eq3}
and \eqref{eq4}: 
\begin{eqnarray}
E & = & \frac{2\tf^{2}-2\tV-E^{2}\tphi'{}^{2}}{4\tf}\,,\label{eq:H-sim}\\
\tf' & = & \frac{E\tphi'{}^{2}}{2}\,,\label{eq:f-prime}\\
\tphi'' & = & -\frac{\tV_{,\tphi}+EE'\tphi'+3E^{2}\tphi'}{E^{2}}\,.\label{eq:ddot-phi-sim}
\end{eqnarray}
The slow-roll parameters can be defined in terms of $E$ as 
\begin{eqnarray}
\epsilon & \equiv & -\frac{\dot{H}}{NH^{2}}=-\frac{E'}{E}=-\frac{(\tf+2E)\tphi'{}^{2}}{2\tf}\,,\label{ep-sim}\\
\epsilon_{1} & \equiv & \frac{\epsilon'}{\epsilon}=\frac{2\tphi''}{\tphi'}-\frac{2E\epsilon}{\tf+2E}-\frac{E^{2}\tphi'{}^{2}}{\tf^{2}+2E\tf}\,,\label{ep1-sim}\\
\eta & \equiv & \frac{\tV_{,\tphi,\tphi}}{\tV}\,.\label{def-eta}
\end{eqnarray}
Inflation requires $\epsilon<1$, while the slow-roll regime additionally
assumes $|\epsilon_{1}|\ll1$. The usual field-space slow-roll conditions
are 
\begin{equation}
\tphi'{}^{2}\ll1\,,\qquad|\tphi''|\ll|\tphi'|\,,\label{slow1}
\end{equation}
which, in cosmic time ($N=1$), correspond at leading slow-roll order
to 
\begin{equation}
\dot{\tphi}^{2}\ll\tV H^{2}_{i}\,,\qquad|\ddot{\tphi}|\ll H|\dot{\tphi}|\,.
\end{equation}
For an expanding universe, $E>0$, Eq.~\eqref{eOfF} implies 
\begin{equation}
2+c\tf>0\,\;\text{for}\,\;\tf<0\,,\quad\;2+c\tf<0\,\;\text{for}\,\;\tf>0\,.\label{fc-con}
\end{equation}
Equation~\eqref{ep-sim} can be expressed in terms of $c$ and $\tilde{f}$
as 
\begin{equation}
\epsilon=-\frac{E(1+c\tf)\tphi'{}^{2}}{\tf(2+c\tf)}\,.
\end{equation}
For $\tilde{f}>0$, $E$ and $\epsilon$ cannot both be positive.
Since a positive $\epsilon$ is required for a conventional graceful
exit, we henceforth restrict attention to $\tilde{f}<0$. In this
branch, $E>0$ and $\epsilon>0$ are compatible provided $c|\tilde{f}|<1$.
At the initial time, $E=1$, and Eq.~\eqref{eOfF} gives 
\begin{equation}
\tf_{i}=\frac{1}{c}\left(-1\pm\sqrt{1-2c}\right)\,.\label{fini}
\end{equation}
Only the plus-sign branch is viable, because it satisfies $1+c\tilde{f}>0$
and $\tilde{f}<0$; the minus-sign branch does not. A real, negative
initial value requires $c<1/2$. Negative values of $c$ can also
yield $E>0$ and $\epsilon>0$, although sufficiently negative $c$
shortens inflation below the duration needed to address the horizon
and flatness problems. For the subsequent analysis, we define 
\begin{align}
\epsilon_{s} & \equiv\frac{1}{2}\tphi'{}^{2}\,,\label{eps-def}\\
\eta_{s} & \equiv\frac{\epsilon_{s}'}{\epsilon_{s}}\,,\label{etas-def}\\
\eta_{s2} & \equiv\frac{\eta_{s}'}{\eta_{s}}=\delta_{s}\,.
\end{align}
The field slow-roll parameters in Eqs.~\eqref{eps-def} and \eqref{etas-def}
are related to the background and potential slow-roll quantities in
Eqs.~\eqref{ep-sim} and \eqref{def-eta} by 
\begin{eqnarray}
\epsilon_{s} & \simeq & \frac{\tf^{2}\epsilon}{\tV-2\tf^{2}}\,,\\
\eta_{s} & = & 4\epsilon-\frac{8\tf^{2}\tV\eta}{3(\tf^{2}-\tV)^{2}}\,,\label{etas-ep}
\end{eqnarray}
where the slow-roll form of the background equations has been used,
in particular 
\begin{equation}
\tphi'\simeq-\frac{\tV_{,\tphi}}{3E^{2}}\,.\label{kg-slow}
\end{equation}
To compare with the standard GR slow-roll expressions, we introduce
\begin{equation}
\epsilon_{v}\equiv\frac{1}{2}\left(\frac{\tV_{,\tphi}}{\tV}\right)^{2}\,.
\end{equation}
This is the standard potential slow-roll parameter entering the GR
expressions for the primordial observables. The variable $\tf$ can
be eliminated from the expressions for the slow-roll parameters by
expressing $\tf$ in terms of $E$ using Eq.~\eqref{eOfF} as 
\begin{equation}
\tf=\frac{1}{c}\left(-1+\sqrt{1-2cE}\right)\,.\label{fofn}
\end{equation}
Substituting this expression into Eq.~\eqref{cSol} gives 
\begin{equation}
\frac{2}{c^{2}}\left(-1+\sqrt{1-2cE}\right)+\frac{2}{c}E\sqrt{1-2cE}+\tV+E^{2}\epsilon_{s}=0\,.\label{eq-nn}
\end{equation}
This algebraic equation determines $E$ in terms of the scalar potential
and kinetic energy. Expanding to first order in $c$ gives 
\begin{equation}
E^{2}=E^{2}_{E}-\frac{2E^{5}_{E}}{\tV}c+{\cal O}(c^{2})\,,\label{e-itter}
\end{equation}
where $E_{E}$ denotes the Einstein-limit solution, 
\begin{equation}
E_{E}=\sqrt{\frac{\tV}{3-\epsilon_{s}}}\simeq\sqrt{\frac{\tV}{3}}\left(1+\frac{\epsilon_{s}}{6}\right)\,.\label{EE}
\end{equation}
Using Eqs.~\eqref{e-itter} and \eqref{EE} in Eq.~\eqref{kg-slow},
we obtain 
\begin{equation}
\epsilon_{s}=\epsilon_{v}+\frac{4c\epsilon_{v}\sqrt{\tV}}{3\sqrt{3}}-\frac{2}{3}\epsilon_{s}\epsilon_{v}\,.\label{eps-ar}
\end{equation}
The slow-roll parameter in Eq.~\eqref{ep-sim} can be written as
\begin{equation}
\epsilon=\epsilon_{v}+\frac{c\epsilon_{v}\sqrt{\tV}}{3\sqrt{3}}-\frac{2}{3}\epsilon_{s}\epsilon_{v}\,.\label{ep-ar}
\end{equation}
For $\eta_{s}$ in Eq.~\eqref{etas-ep}, we can write 
\begin{equation}
\eta_{s}\simeq4\epsilon_{v}-2\eta+\frac{4c}{3\sqrt{3}}(\epsilon_{v}-\eta)\sqrt{\tV}\,.\label{etas-ar}
\end{equation}
Equations~\eqref{eps-ar} and \eqref{ep-ar} show that the $c$-dependent
corrections are suppressed by the product of the deformation parameter
and a slow-roll quantity. The terms proportional to $\epsilon_{s}\epsilon_{v}$
are formally of higher slow-roll order and have been retained as generated
by the iterative expansion.

\subsection{Primordial power spectra}

\subsubsection{Curvature perturbation}

Since the metric curvature perturbation is constrained to vanish in
the spatial gauge used above, the gauge-invariant comoving curvature
perturbation is carried entirely by the scalar-field fluctuation,
\begin{equation}
\delta\phi=-\frac{\dot{\phi}}{NH}\,\mathcal{R}=-\Mpl\,\tilde{\phi}'\,\mathcal{R}.
\end{equation}
Introducing conformal time through 
\begin{equation}
{\rm d}\tau=\frac{N}{a}\,{\rm d}t
\end{equation}
and defining 
\begin{equation}
Q_{s}\equiv\Mpl^{2}\epsilon_{s},
\end{equation}
the quadratic action takes the form below. In this subsection, primes
on perturbations denote derivatives with respect to $\tau$, whereas
$\epsilon_{s}$, $\eta_{s}$, and $\eta_{s2}$ retain their e-fold
definitions. 
\begin{align}
\mathcal{S}_{\mathcal{R}}=\int{\rm d}\tau\,{\rm d}^{3}x\,a^{2}Q_{s}\Bigg[ & (\mathcal{R}')^{2}-(\partial_{i}\mathcal{R})(\partial^{i}\mathcal{R})\nonumber \\
 & +\frac{a^{2}}{2}\Biggl(H^{2}\eta_{s}[\epsilon-3-\eta_{s2}]-2C_{\phi}\nonumber \\
 & \hspace{5em}-\frac{H^{2}}{2}\eta^{2}_{s}\Biggr)\mathcal{R}^{2}\Bigg].\label{eq:curvature-action}
\end{align}
Thus the scalar sound speed is unity and the no-ghost condition is
simply $Q_{s}>0$, or equivalently $\epsilon_{s}>0$.

For a Fourier mode, the exact equation following from Eq.~\eqref{eq:curvature-action}
can be written as 
\begin{equation}
\begin{aligned}0={} & \mathcal{R}_{k}''+aH(2+\eta_{s})\mathcal{R}_{k}'+k^{2}\mathcal{R}_{k}\\
 & +a^{2}\left[C_{\phi}+\frac{H^{2}}{4}\eta^{2}_{s}-\frac{H^{2}}{2}(\epsilon-3-\eta_{s2})\eta_{s}\right]\mathcal{R}_{k}\,.
\end{aligned}
\label{eq:ddotR}
\end{equation}
With the canonical variable 
\begin{equation}
v_{k}=z_{s}\mathcal{R}_{k},\qquad z_{s}\equiv a\sqrt{2Q_{s}},
\end{equation}
and $C_{\phi}=H^{2}\bar{C}_{\phi}$, the mode equation is 
\begin{equation}
v_{k}''+\left[k^{2}-a^{2}H^{2}\left(2-\epsilon-\bar{C}_{\phi}\right)\right]v_{k}=0.
\end{equation}
For slow roll, $\bar{C}_{\phi}=\mathcal{O}(\epsilon)$ and $aH\simeq-1/\tau$,
so the leading equation is 
\begin{equation}
v_{k}''+\left(k^{2}-\frac{2}{\tau^{2}}\right)v_{k}=0.
\end{equation}
The Bunch--Davies solution gives the scalar power spectrum at scalar
horizon exit, $k=aH$, as 
\begin{equation}
\mathcal{P}_{s}=\frac{H^{2}_{*}}{8\pi^{2}\Mpl^{2}\epsilon_{s*}}.\label{eq:scalar-power-spectrum}
\end{equation}
Here and below a star denotes evaluation at the relevant horizon-crossing
time. To first order in slow roll, 
\begin{equation}
n_{s}-1=\frac{{\rm d}\ln\mathcal{P}_{s}}{{\rm d}\ln k}=-\frac{2\epsilon+\eta_{s}}{1-\epsilon}\simeq-2\epsilon-\eta_{s}.\label{ns}
\end{equation}

\subsubsection{Tensor perturbations}

For either tensor polarization, the quadratic action of the toy model
is 
\begin{equation}
\mathcal{S}_{T}=\int{\rm d}t\,{\rm d}^{3}x\,Na^{3}Q_{T}\left[\frac{\dot{h}^{2}}{N^{2}}-c^{2}_{T}\frac{(\partial_{i}h)(\partial^{i}h)}{a^{2}}\right],
\end{equation}
where 
\begin{equation}
Q_{T}=\frac{\Mpl^{2}}{8c^{2}_{T}},\qquad c^{2}_{T}=1+\frac{3}{2}\,\tilde{f}\,(\tilde{\xi}_{2}+\tilde{\xi}_{3}).\label{eq:tensor-sound-speed}
\end{equation}
The tensor no-ghost and gradient conditions are therefore 
\begin{equation}
Q_{T}>0,\qquad c^{2}_{T}>0.
\end{equation}
Defining the canonical variable $v_{T,k}=a\sqrt{2Q_{T}}\,h_{k}$,
the leading slow-roll equation is 
\begin{equation}
v_{T,k}''+\left(c^{2}_{T}k^{2}-\frac{2}{\tau^{2}}\right)v_{T,k}=0,
\end{equation}
where the slow variation of $c_{T}$ and $Q_{T}$ has been neglected
at this order. Summing over the two polarizations, the tensor spectrum
at tensor horizon exit, $c_{T}k=aH$, is 
\begin{equation}
\mathcal{P}_{T}=\frac{2H^{2}_{T*}}{\pi^{2}\Mpl^{2}c_{T*}}.\label{eq:tensor-power-spectrum}
\end{equation}
Neglecting the slow-roll difference between scalar and tensor crossing
times gives 
\begin{equation}
r\equiv\frac{\mathcal{P}_{T}}{\mathcal{P}_{s}}=\frac{16\epsilon_{s}}{c_{T}}.\label{r}
\end{equation}
Thus $c_{T}>1$ suppresses the tensor-to-scalar ratio. At the same
time, $Q_{T}=\Mpl^{2}/(8c^{2}_{T})$ decreases as $c_{T}$ grows.
To clarify the implication of this behavior, consider the perturbative
expansion $S=S^{(2)}+S^{(3)}+S^{(4)}+\cdots$. The strong-coupling
scale is the energy at which interactions among canonically normalized
fluctuations become of order unity, so that the higher-order terms
can no longer be treated as small corrections to $S^{(2)}$. On a
time-dependent background with broken Lorentz invariance, this scale
can itself be time dependent, and the energy and momentum cutoffs
need not coincide.

At fixed background time and suppressing powers of the scale factor
and the detailed derivative structure,
let $h_{c}\equiv\sqrt{2Q_{T}}\,h$ denote the locally normalized tensor
fluctuation. Consider schematically an interaction
$\lambda_{n}\mathcal{O}_{n}[h]$ containing $n$ powers of the tensor
perturbation. In terms of the canonical field it becomes
\begin{equation}
\lambda_{n}\mathcal{O}_{n}[h]
=\frac{\lambda_{n}}{(2Q_{T})^{n/2}}
\mathcal{O}_{n}[h_{c}]\,.
\end{equation}
Thus, a small $Q_T$ enhances canonically normalized tensor interactions
unless their coefficients carry compensating powers of $Q_T$. In the
present model,
\begin{equation}
\sqrt{2Q_T}=\frac{\Mpl}{2c_T}\,,
\end{equation}
so a parametrically large $c_T$ reduces the tensor canonical-normalization
coefficient. At fixed coefficients of the nonlinear operators, this
enhances the corresponding canonically normalized interactions and
may lower the perturbative cutoff. This argument does not determine
the actual strong-coupling scale, because the coefficients of the
cubic and higher-order operators can themselves depend on $c_T$.
Establishing that scale requires the interaction action and cannot
be inferred from $S^{(2)}$ alone.

By a possible ``loss of perturbative control'' we mean that ratios
such as $S^{(3)}/S^{(2)}$, higher-order contributions, or loop
corrections may cease to be small at the scales used to compute the
primordial spectra. For those predictions to be reliable, the
characteristic frequencies of order $H$ and the physical momenta
probed around horizon crossing must remain parametrically below the
corresponding energy and momentum cutoffs. If this hierarchy is lost,
predictions based only on the quadratic action are not controlled.
This is a limitation of the perturbative description, rather than by
itself a proof of an instability or of an inconsistency of the full
theory.

\subsection{Observational constraints}

For definiteness, we adopt the rounded bounds $n_{s}=0.965\pm0.004$
and $r<0.06$, based on the Planck 2018 analysis and its combination
with the BICEP2/Keck Array BK15 data \cite{Planck:2018jri}. These
values serve as illustrative benchmarks rather than as an updated
likelihood analysis.

For small $c$, Eqs.~\eqref{eps-ar}, \eqref{etas-ar}, \eqref{ns},
and \eqref{r} give 
\begin{align}
n_{s}-1 & \simeq-6\epsilon_{v}+2\eta+\frac{2\sqrt{\tV}}{3\sqrt{3}}c(2\eta-3\epsilon_{v})+4\epsilon_{s}\epsilon_{v}-\frac{2}{3}\epsilon_{s}\eta,\\
r & \simeq\frac{16\epsilon_{v}}{c_{T}}\left(1+\frac{4c\sqrt{\tV}}{3\sqrt{3}}-\frac{2}{3}\epsilon_{s}\right).
\end{align}
These expressions recover the GR slow-roll results when the deformation
is removed and $c_{T}\to1$.

\begin{figure}
\IfFileExists{ns.pdf}{\includegraphics[width=0.8\columnwidth]{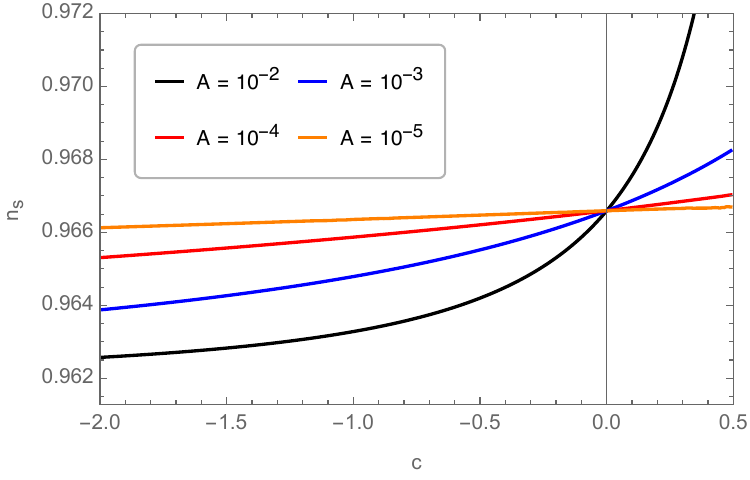}}{%
\fcolorbox{black}{white}{\parbox[c][3cm]{0.45\columnwidth}{%
\centering Missing figure: ns%
}}}
\IfFileExists{rcT.pdf}{\includegraphics[width=0.8\columnwidth]{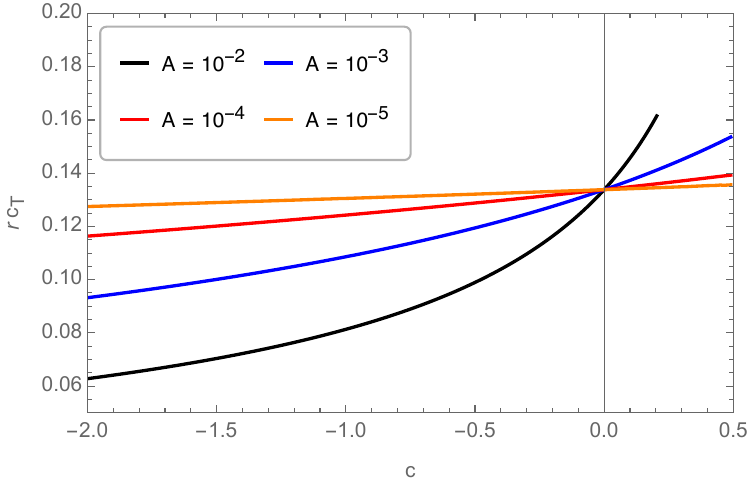}}{%
\fcolorbox{black}{white}{\parbox[c][3cm]{0.45\columnwidth}{%
\centering Missing figure: rcT%
}}} \caption{The scalar spectral index $n_{s}$ and the combination $rc_{T}=16\epsilon_{s}$
as functions of $c$ for several values of $A$. Each curve terminates
when $n_{s}$ leaves the adopted interval $0.965\pm0.004$.}
\label{fig:ns-r} 
\end{figure}

To compute observational quantities for the full range of $c$ considered
here, we solve Eqs.~\eqref{eq:f-prime} and \eqref{eq:ddot-phi-sim}
numerically. We choose the potential of the scalar field in the form
\begin{equation}
\tV=A\tphi^{2}\,,
\end{equation}
where $A$ is constant. The initial condition for $\tf$ is set from
Eq.~\eqref{fini}, while the initial conditions for $\tphi$ and
$\tphi'$ are computed from Eqs.~\eqref{cSol} and \eqref{ep-sim}
by setting $\epsilon=10^{-3}$ at the initial time. The initial value
of $\epsilon$ has no effect on the observable predictions provided
that it is sufficiently small. The observables are evaluated for modes
that cross the Hubble radius 60 e-folds before the end of inflation.
Plots of $n_{s}$ and the combination $rc_{T}=16\epsilon_{s}$ for
various values of $c$ and $A$ are shown in Fig.~\ref{fig:ns-r}.

\begin{figure}
\IfFileExists{regionAC.pdf}{\includegraphics[width=0.8\columnwidth]{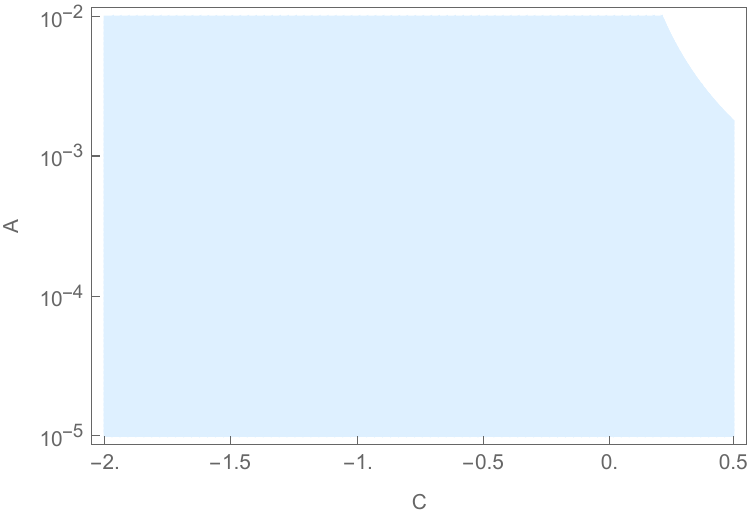}}{%
\fcolorbox{black}{white}{\parbox[c][3cm]{0.45\columnwidth}{%
\centering Missing figure: regionAC%
}}} \IfFileExists{c-cT.pdf}{\includegraphics[width=0.8\columnwidth]{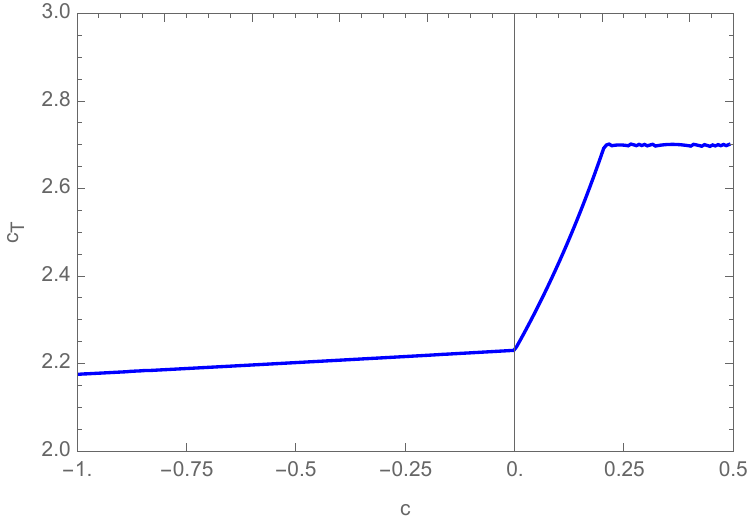}}{%
\fcolorbox{black}{white}{\parbox[c][3cm]{0.45\columnwidth}{%
\centering Missing figure: c-cT%
}}} \caption{Left: region in the $(c,A)$ plane satisfying the illustrative range
$n_{s}=0.965\pm0.004$. Right: minimum $c_{T}$ required to ensure
$r\protect\leq0.06$ under the observational bound adopted here.}
\label{fig:region-sound-speed} 
\end{figure}

The left panel shows that $n_{s}$ decreases as either $c$ or $A$
is reduced and approaches the Einstein-gravity prediction when both
parameters are sufficiently small. The combination $rc_{T}$ also
decreases toward smaller $c$, reflecting the corresponding reduction
of $\epsilon_{s}$. Figure~\ref{fig:region-sound-speed} displays
the region in the $(c,A)$ plane compatible with the illustrative
scalar-tilt range adopted here. The right panel shows, at each fixed
$c$, the minimum $c_{T}$ that guarantees the adopted upper bound
on $r$ over the entire allowed range of $A$. It is obtained by first
maximizing $\epsilon_{s}$ with respect to $A$ and then using that
maximum in Eq.~\eqref{r}. For a particular pair $(c,A)$, the required
value of $c_{T}$ may be lower than the envelope shown in the right
panel. The plotted curve is instead the conservative minimum that
guarantees the adopted upper bound on $r$ throughout the allowed
range of $A$ at fixed $c$.

\section{Conclusions}

\label{sec:conclusions}

We have developed a modified four-constraint MMG construction designed
to yield a closed and predictive FLRW cosmology. In the original implementation
of Ref.~\cite{Yao:2023qjd}, the homogeneous Lagrange multipliers
are not fully determined by the background equations, although they
enter the quadratic tensor action. The propagation of gravitational
waves consequently depends on arbitrary functions that are not part
of the physical initial data.

Our modification replaces the direct multiplier coupling by a coupling
to the spatial Laplacian of each multiplier. Variation with respect
to the multipliers gives the elliptic constraints $D^{2}(\mathcal{Q}^{I}-\mathcal{P}^{I})=0$.
For inhomogeneous modes and standard boundary conditions, these impose
the same local constraints as the original construction, while the
homogeneous multiplier sector drops out on FLRW. The resulting tensor
action is completely determined by the Hamiltonian and the background
solution. Its dispersion relation is 
\begin{equation}
\omega^{2}=\frac{D_{T}}{Q_{T}}\frac{k^{4}}{a^{4}}+\frac{C_{T}}{Q_{T}}\frac{k^{2}}{a^{2}},
\end{equation}
with $Q_{T}>0$ required to avoid tensor ghosts. We also verified
that no gravitational vector or scalar mode propagates on the generic
FLRW branch for which the constraint reduction is nondegenerate, for
$k\neq0$.

After coupling a canonical scalar field, the scalar sector contains
a single propagating mode, represented either by $\delta\phi$ or
by the comoving curvature perturbation $\mathcal{R}$. Its quadratic
action has a canonical kinetic term and unit sound speed. We distinguished
the generic case with a nonvanishing lapse Hessian, in which preservation
of the algebraic lapse constraint determines $\dot N$, from the subclass
$\tilde{\mathscr{H}}^{{\rm CH}}=N\mathcal{F}$. In the latter subclass
the homogeneous equations depend on the lapse
only through $N\,{\rm d}t$, so the lapse can be absorbed into a time
redefinition at the background level.

For the cubic-momentum inflationary model, the $k^{4}$ tensor term
vanishes and 
\begin{equation}
c^{2}_{T}=1+\frac{3}{2}\tilde{f}(\tilde{\xi}_{2}+\tilde{\xi}_{3}),\qquad Q_{T}=\frac{\Mpl^{2}}{8c^{2}_{T}}.
\end{equation}
The background deformation is controlled by $c=-3(\tilde{\xi}_{3}+9\tilde{\xi}_{1}+3\tilde{\xi}_{2})$,
with the GR background limit obtained for $c\to0$. Reality of the
viable expanding branch requires $c<1/2$, while the duration
of inflation further restricts sufficiently negative values. At leading
order in slow roll, 
\begin{equation}
n_{s}-1\simeq-2\epsilon-\eta_{s},\qquad r=\frac{16\epsilon_{s}}{c_{T}}.
\end{equation}
Our numerical analysis for a quadratic potential shows that increasing
the deformation $c$ generally raises both $n_{s}$ and $rc_{T}=16\epsilon_{s}$.
Within the observational bounds adopted in this work, the viable region
requires $c_{T}>1$ in order to suppress $r$. This observational
benefit must be considered together with the decrease of $Q_T$ at
large $c_T$: canonically normalized nonlinear interactions may then
be enhanced, although the actual perturbative cutoff can only be
established from the cubic and higher-order actions.

Several issues remain open. These include computing the strong-coupling
scale, studying the transition to a late-time GR regime, and confronting
the model with current joint constraints on scalar and tensor primordial
spectra. 


\appendix

\section{Quadratic scalar action and constraint reduction}

\label{appendix1}

For completeness, we summarize the structure of the scalar reduction.
In the gauge $E=0$, the perturbations are $\{\alpha,\chi,\zeta,\zeta_{\pi},E_{\pi},\delta\mu_{I}\}$
in vacuum and additionally $\delta\phi$ in the matter-coupled theory.
The gravitational quadratic action can be organized as 
\begin{equation}
S^{(2)}_{g}=\int{\rm d}t\,{\rm d}^{3}x\,\mathcal{L}^{(2)}_{g},\label{act-s}
\end{equation}
with the schematic form 
\begin{align}
\frac{\mathcal{L}^{(2)}_{g}}{Na^{3}}={} & A_{1}\zeta\dot{\zeta}+A_{2}\alpha\dot{\zeta}+A_{3}(\partial^{2}E_{\pi})\dot{\zeta}+A_{4}\zeta_{\pi}\dot{\zeta}\nonumber \\
 & +B_{1}\zeta^{2}+B_{2}\zeta\partial^{2}\chi+B_{3}\zeta\zeta_{\pi}+B_{4}\zeta\partial^{2}E_{\pi}\nonumber \\
 & +C_{1}\alpha\partial^{2}\delta\mu_{1}+C_{2}\alpha\partial^{2}\delta\mu_{2}+C_{3}\alpha\partial^{2}\delta\mu_{3}\nonumber \\
 & +D_{1}(\partial^{2}\zeta)(\partial^{2}\delta\mu_{1})+D_{2}(\partial^{2}\zeta)(\partial^{2}\delta\mu_{2})+\mathcal{L}_{{\rm alg}},\label{lan-s}
\end{align}
where $\mathcal{L}_{{\rm alg}}$ contains terms algebraic in $\alpha$,
$\chi$, $\zeta_{\pi}$, and $E_{\pi}$. The coefficients are functions
of the homogeneous background and derivatives of $\tilde{\mathscr{H}}^{{\rm CH}}$.
Their explicit unreduced expressions are not needed for the degree-of-freedom
count.

In Fourier space, variation with respect to the multiplier perturbations
gives, on the generic branch, 
\begin{equation}
k^{2}\mathcal{P}^{3}{}_{,N}\alpha=0,\qquad k^{4}\frac{f}{a}\zeta=0.
\end{equation}
Thus $\alpha=\zeta=0$ for $k\neq0$, $f\neq0$, and $\mathcal{P}^{3}{}_{,N}\neq0$.
The shift equation fixes 
\begin{equation}
E_{\pi}=\frac{a^{2}}{k^{2}}\zeta_{\pi}
\end{equation}
in vacuum. The remaining algebraic equation sets $\zeta_{\pi}=0$
whenever its coefficient is nonzero, proving the absence of an inhomogeneous
gravitational scalar mode on this branch.

With a canonical scalar field, the full quadratic action is 
\begin{equation}
S^{(2)}_{S}=S^{(2)}_{g}+S^{(2)}_{\phi}.\label{act-sc-phi}
\end{equation}
After imposing $\alpha=\zeta=0$, the matter contribution relevant
for the remaining constraints is 
\begin{align}
S^{(2)}_{\phi}\big|_{\alpha=\zeta=0}={} & \frac{1}{2}\int{\rm d}t\,{\rm d}^{3}x\,Na^{3}\Bigg[\nonumber \\
 & \frac{\dot{\delta\phi}^{\,2}}{N^{2}}-\frac{(\partial_{i}\delta\phi)(\partial^{i}\delta\phi)}{a^{2}}\nonumber \\
 & -V_{,\phi\phi}\delta\phi^{2}-\frac{2\dot{\phi}}{N^{2}a^{2}}(\partial_{i}\chi)(\partial^{i}\delta\phi)\Bigg].
\end{align}
The shift constraint becomes 
\begin{equation}
E_{\pi}=\frac{a^{2}}{k^{2}}\zeta_{\pi}-\frac{a^{2}}{4fk^{2}}\frac{\dot{\phi}}{N}\delta\phi.
\end{equation}
Eliminating $E_{\pi}$ and $\zeta_{\pi}$ then gives the reduced scalar
action quoted in Eq.~\eqref{act-scalar}. The reduction assumes that
the relevant algebraic denominator $E_{\phi}$ is nonzero; if it vanishes,
the system belongs to a separate degenerate branch and must be reconsidered
before drawing conclusions about its degree-of-freedom content.

 \bibliographystyle{apsrev4-2}
\bibliography{draft}

\end{document}